\documentclass{article}
\usepackage{geometry}       
\usepackage{authblk}
\usepackage{amsmath}
\usepackage{upgreek}
\usepackage{booktabs}
\usepackage{multirow}
\usepackage{rotating}
\usepackage{tabularx}
\usepackage{array}
\usepackage{makecell}
\usepackage{placeins}
\usepackage{bm}
\newcolumntype{Y}{>{\raggedright\arraybackslash}X}
\newcolumntype{Z}{>{\centering\arraybackslash}X}
\usepackage{lineno}
\usepackage{graphicx}
\usepackage{subcaption}

\usepackage[numbers]{natbib}
\usepackage[colorlinks=true,allcolors=blue]{hyperref}
\usepackage{doi}

\title{Robust watt-level continuous-wave deep-ultraviolet lasers near 230~nm}

\author[2,5$\dagger$]{J. Cai}
\author[1,6$\dagger$]{M. Stoepper}
\author[5]{P. Agarwal}
\author[2]{L. Pal\'{a}nki}
\author[2]{C. Alarc\'{o}n-Robledo}
\author[2]{W. W. W. Liu}
\author[5]{P. Kukreja}
\author[8,9]{R. Ferstl}
\author[8,9]{H. Foltas}
\author[7]{L. M\"oller}
\author[2]{D. J. Brown}
\author[2]{C. J. H. Rich}
\author[3]{M. Chiarotti}
\author[10]{T. Uusitalo}
\author[5]{R. Thomas}
\author[4]{J. Domarkas}
\author[10]{M. Guina}
\author[10]{J.-P. Penttinen}
\author[7]{S. Stellmer}
\author[3]{N. Poli}
\author[5]{S. C. Wright}
\author[1]{S. Hannig}
\author[2,5*]{S. Truppe}

\affil[1]{Agile Optic GmbH, Krähenfeld 11, 38110 Braunschweig, Germany}
\affil[2]{Centre for Cold Matter, Blackett Laboratory, Imperial College London, London, SW7 2AZ, United Kingdom}
\affil[3]{Dipartimento di Fisica e Astronomia and LENS, Universit\`a degli Studi di Firenze, INFN Sezione di Firenze, Via Sansone 1, 50019 Sesto Fiorentino, Italy}
\affil[4]{EKSMA Optics UAB, Dvarcioniu str. 2B, LT-10233 Vilnius, Lithuania}
\affil[5]{Fritz-Haber-Institut der Max-Planck-Gesellschaft, Faradayweg 4-6, 14195 Berlin, Germany}
\affil[6]{Leibniz Universit\"at Hannover, Welfengarten 1, 30167 Hannover, Germany}
\affil[7]{University of Bonn, Physics Institute, Nussallee 12, 53115 Bonn, Germany}
\affil[8]{University of Vienna, Faculty of Physics, Quantum Science, Boltzmanngasse 5, 1090 Vienna, Austria}
\affil[9]{University of Vienna, Vienna Doctoral School in Physics, Boltzmanngasse 5, 1090 Vienna, Austria}
\affil[10]{Vexlum Ltd., Kauhakorvenkatu 53 B, 33710 Tampere, Finland}

\affil[$\dagger$]{These authors contributed equally to this work.}
\affil[*]{\href{mailto:s.truppe@imperial.ac.uk}{s.truppe@imperial.ac.uk}}
\begin{document}
\maketitle
\begin{abstract}
Continuous-wave (CW) deep-ultraviolet (DUV) lasers near 230~nm enable laser cooling of AlF, Cd, and Zn, but second-harmonic generation below 237~nm relies in practice on beta-barium borate (BBO), whose walk-off and UV-induced degradation hinder sustained operation. We demonstrate compact, affordable VECSEL-based systems informed by four years of operating 14 DUV cavities in 12 laser systems across six European laboratories. External LBO cavities produce nearly 4~W at 463~nm with 94\% cavity efficiency. We compare spherically and elliptically focused Brewster-cut BBO cavities with a normal-incidence AR-coated design. The AR-coated cavity delivers the highest power and efficiency, reaching 1.0~W at 51\% cavity and 44\% external efficiency; the spherical Brewster cavity reaches 700~mW and maintains constant circulating power over 70~h, while elliptical focusing reduces peak intensity sixfold and improves beam quality, albeit with greater alignment sensitivity. Collaboration-designed DUV optics, AlF spectroscopy, and Cd trapping validate the system.
\end{abstract}

\section{Introduction}
Continuous-wave (CW) laser sources in the deep ultraviolet (DUV) enable laser cooling and precision spectroscopy of new atomic and molecular species, alongside applications in photochemistry and semiconductor inspection. In the 210--240~nm range, cadmium has a strong dipole-allowed transition for laser cooling near 229~nm~\cite{brickman_magneto-optical_2007,PadillaCd2025,Tinsley2024,Tinsley2021}, zinc has a corresponding transition near 214~nm~\cite{moller2025}, and aluminium monofluoride (AlF) possesses highly closed cycling transitions near 227.5~nm, making it well suited to laser cooling, magneto-optical trapping, precision measurement, and quantum simulation~\cite{Truppe2019,Hofsass2021,Padilla2025}. Nearby wavelengths are also relevant to trapped-ion qubits~\cite{Burd2016,Burd2023}, Rydberg excitation~\cite{feldker_rydberg_2015,zhang_submicrosecond_2020}, hydrogen and muonium spectroscopy~\cite{parthey_improved_2011,crivelli_mu-mass_2018,yzombard_1s3s_2023}, next-generation matter-wave interferometry~\cite{Pedalino2026}, resonance-enhanced Raman scattering~\cite{asher_uv_1993}, and high-resolution materials diagnostics~\cite{stokowski_wafer_1998}. These applications share a common set of requirements: high power combined with narrow linewidth, good spatial mode quality, and stable long-term operation.

Meeting these requirements remains technically challenging. Direct solid-state UV emitters are at an early stage of development~\cite{amano_2020}, and frequency-converted systems based on Ti:sapphire lasers (4--5~W in the 910--926~nm range) deliver high fundamental powers but remain expensive, complex, and large. We have built such systems ourselves, most recently a quadrupled Ti:sapphire chain that produces 130~mW of CW light at 214~nm~\cite{moller2025Laser} and enabled magneto-optical trapping of zinc~\cite{moller2025}; they are used here as the performance benchmark against which the more compact, less complex, and lower-cost vertical-external-cavity surface-emitting laser (VECSEL) architectures are assessed. External-cavity diode lasers with tapered amplifiers (3--3.5~W) provide another compact route, but the poor spatial mode of the tapered amplifier limits the usable power and the frequency-doubling efficiency.

More recently, frequency-doubled thulium-doped fiber amplifiers have become available with single-frequency powers of 7--10~W in the 910--926~nm range, exceeding that of a Ti:sapphire laser. This power, however, is obtained at a considerable practical cost. The seed laser, the multi-stage amplifier chain, its pump diodes, and the frequency-doubling stage together form a complex system with a large physical footprint and a high cost, and the thermal load of the amplifier chain requires substantial water-cooling infrastructure.

Beyond the choice of fundamental laser source, frequency conversion itself becomes rapidly more demanding towards shorter wavelengths. At 266~nm, robust low-walk-off crystals such as cesium lithium borate (CLBO) deliver multi-watt CW power and remain usable down to about 240~nm~\cite{Shaw2021,Burkley2019}. CLBO's $\sim$237~nm limit, however, applies to second-harmonic generation. Sum-frequency mixing in CLBO phase-matches to much shorter wavelengths and is the established route to 193~nm, including in the CW regime~\cite{Sakuma2011}, but it requires a second single-frequency source at a different wavelength and one further conversion stage. For a compact system using a single fundamental laser, beta-barium borate (BBO) therefore remains the only practical crystal below 237~nm, and its larger walk-off and stronger UV absorption have so far limited CW output below 230~nm to sub-watt levels; the highest reported CW powers are 0.56~W at 229~nm~\cite{Kaneda2016} and 0.46~W at 213~nm~\cite{kaneda_scalable_2019}.

These demonstrations establish the power that is achievable, but rarely the sustained, long-term operation that applications require. Narrowing the gap between demonstrated power and reliable long-term operation motivates the work reported here.

Here we demonstrate compact laser architectures based on VECSELs~\cite{Guina2017} for generating CW DUV radiation near 230~nm, covering the main AlF cooling wavelengths (227.5~nm, 231.7~nm) and the Cd cooling wavelength (229~nm). The systems combine 910--926~nm VECSELs, external or intracavity second-harmonic generation (SHG) to blue wavelengths, and resonant SHG in BBO to reach the DUV. The normal-incidence AR-coated BBO cavity reported here delivers up to 1.0~W near 230~nm, which to our knowledge is the highest continuous-wave power reported from direct second-harmonic generation below 237~nm. Compared with previous VECSEL-based DUV sources~\cite{Kaneda2016}, the present work emphasizes long-term operation and damage mitigation through tailored BBO focusing geometries, improved thermal management, and DUV-compatible optical components, while reducing the cost, footprint, and operational complexity of conventional conversion chains.

Additionally, we characterize low-loss mirrors, thin-film polarizing beam splitters, and anti-reflection-coated vacuum windows developed for this wavelength range. We validate them by using them to operate a magneto-optical trap (MOT) of $^{112}$Cd on the $^1S_0\rightarrow{}^1P_1$ transition at 228.9~nm~\cite{PadillaCd2025}. A DUV MOT is a demanding test of the complete optical chain, since it requires high delivered power, a clean wavefront, and transmission that remains stable over the hours of a typical experiment.

Although the experimental results presented below focus on VECSEL sources, the conclusions draw on the broader UVQuanT program, within which 14 CW DUV cavities have been operated as part of 12 laser systems at six laboratories across Europe over the past four years, spanning 214~nm~\cite{moller2025Laser} to 232~nm and requiring only minimal servicing.

These systems span Ti:sapphire and VECSEL fundamental sources, external and intracavity blue generation, Brewster-cut and normal-incidence anti-reflection (AR)-coated BBO, and spherical and elliptical focusing geometries, and were operated by different groups for laser cooling and nanocluster interferometry. This allows us to distinguish robust design principles from site-specific implementation details; Table~\ref{tab:duv_systems_overview} summarizes the systems and their representative operating conditions.

The paper is organized as follows. We first describe the VECSEL sources and the near-infrared (NIR)-to-blue conversion stages, including external lithium triborate (LBO) enhancement cavities. We then discuss the BBO cavity designs for DUV generation, with emphasis on large spherical and elliptical focusing geometries that reduce the bulk optical dose and improve the DUV mode profile while maintaining efficient conversion, before turning to the observed damage mechanisms and mitigation strategies for long-term CW operation near 230~nm. Finally, we characterize DUV-compatible mirrors, polarizing optics, and vacuum windows, and demonstrate the system by recording AlF molecular-beam spectra on the $\mathrm{A}^1\Pi(v'=0)\leftarrow\mathrm{X}^1\Sigma^+(v''=0)$ transition and by trapping $^{112}$Cd atoms in a magneto-optical trap at 228.9~nm.

\section{System overview}
\begin{figure*}[!htbp]
\centering
\includegraphics[width=\textwidth]{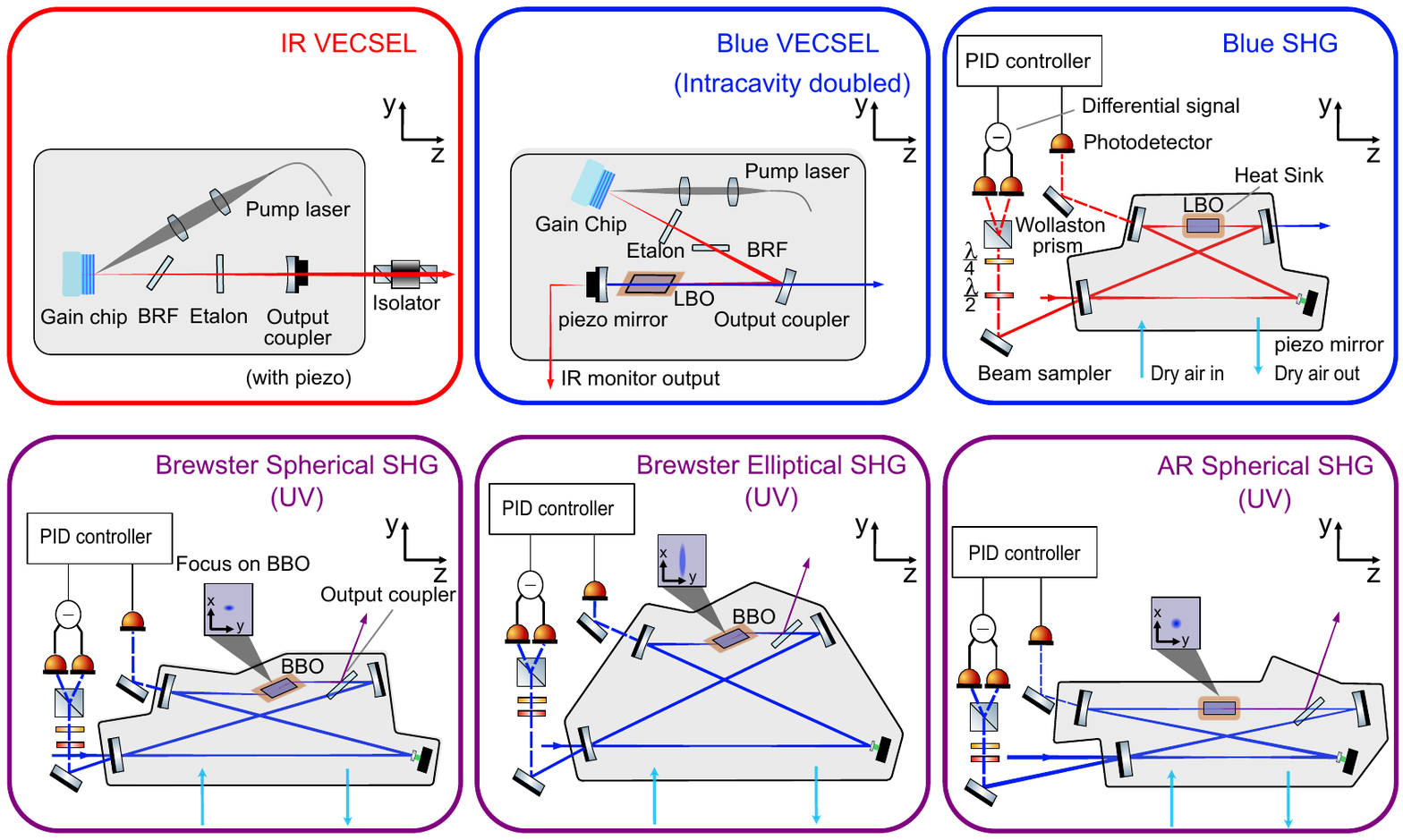}
\caption{Building blocks of the laser systems used to generate deep-ultraviolet (DUV) laser light near 230~nm. Top row: near-infrared (NIR) VECSEL, intracavity frequency-doubled blue VECSEL, and external LBO enhancement cavity for second-harmonic generation (SHG) of blue light (near 460~nm). Bottom row: the three resonant BBO cavity variants used for DUV generation, employing spherical focusing in Brewster-cut BBO, elliptical focusing in Brewster-cut BBO, and spherical focusing in normal-incidence anti-reflection (AR)-coated BBO. In all DUV cavities a dichroic mirror separates the DUV light from the circulating blue field. Insets indicate the fundamental focal spot in the BBO crystal.
}
\label{fig:lasersystem}
\end{figure*}

The majority of our laser systems are based on optically pumped semiconductor VECSELs that provide tunable, narrowband, near-diffraction-limited radiation in the 910--926~nm range. Figure~\ref{fig:lasersystem} summarizes the building blocks of the laser systems used in this work: two VECSEL-based routes to blue light and three DUV cavity variants.

NIR VECSELs produce 910--926~nm light in linear I-cavity geometries (Vexlum Valo SF), followed by external LBO enhancement cavities (Agile Optic) that generate 455--463~nm light. This blue light is then frequency-doubled to the DUV in resonant BBO cavities. Alternatively, an intracavity-doubled VECSEL (Vexlum Valo SHG) uses LBO within its V-cavity to generate 455--463~nm light directly, before the same final BBO stage. The DUV stage itself is implemented in three variants: Brewster-cut BBO with spherical focusing, Brewster-cut BBO with elliptical focusing, and normal-incidence AR-coated BBO with spherical focusing.

These architectures are designed to address the main AlF cooling wavelength near 227.5~nm, the AlF repumping wavelength near 231.7~nm, and the wavelength of the Cd $^1S_0 \rightarrow {}^1P_1$ transition near 229~nm. They also allow a direct comparison of external and intracavity blue generation as pump sources for resonant DUV SHG.

\section{VECSEL platforms}

\subsection{VECSEL gain chips and cavity architectures}
The semiconductor gain chips are grown by solid-source molecular beam epitaxy and consist of multiple gallium indium arsenide (GaInAs) quantum wells embedded in gallium arsenide (GaAs) barriers, centered within a microcavity defined by an aluminium arsenide/gallium arsenide (AlAs/GaAs) distributed Bragg reflector~\cite{Guina2017}. The quantum wells are arranged in a resonant periodic gain configuration, with their centers aligned to the antinodes of the standing wave. Each structure incorporates 12--14 strain-compensated quantum wells. For efficient heat extraction, the chips are flip-chip bonded to synthetic diamond heat spreaders after substrate removal and mounted on water-cooled copper heat sinks.

All VECSELs are pumped by integrated multimode fiber-coupled diodes operating at 800--815~nm, delivering up to 25~W of optical pump power. The pump beam is focused to a spot of $\approx300~\upmu$m diameter on the gain chip. Single-transverse-mode operation is maintained by matching the pump spot to the cavity's fundamental Gaussian (TEM$_{00}$) mode at the gain mirror.

The NIR lasers use linear I-cavities formed by the gain mirror and a curved output coupler. Spectral selection is provided by a birefringent filter and a temperature-stabilized yttrium aluminium garnet (YAG) etalon, while a piezo-mounted output coupler enables fine frequency tuning. The output-coupler transmission is 1--2\%, optimized for the available gain and target wavelength. The visible lasers use a V-shaped cavity in which an LBO crystal placed at an intracavity waist generates 455~nm light directly~\cite{Hill2022}. The folding mirror is highly reflective at 910~nm and transmissive at 455~nm, allowing extraction of the frequency-doubled output.

\subsection{Output power and spatial mode quality}
The VECSELs produce up to 2.1~W at 910~nm and up to 5.4~W at 926~nm; the intracavity-doubled VECSEL delivers up to 1.5~W at 455~nm.

\begin{figure}
    \centering
    \includegraphics[width=\columnwidth]{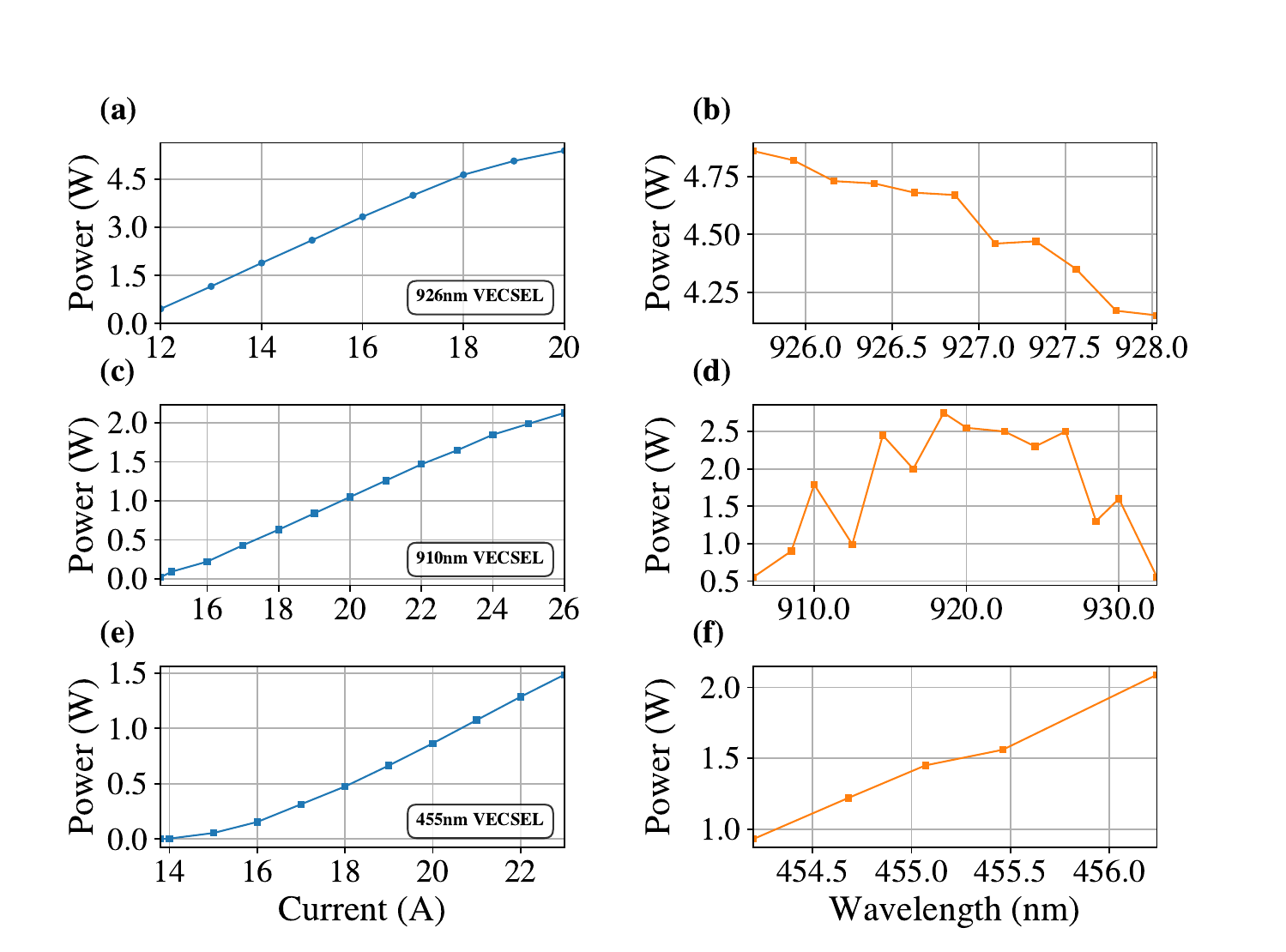}
    \caption{Output power and wavelength tuning of the VECSEL systems. Panels (a,b): 926~nm VECSEL; panels (c,d): 910~nm VECSEL; panels (e,f): intracavity-doubled 455~nm VECSEL. In panels (a,c,e), output power is plotted against pump-diode current; in panels (b,d,f), output power is plotted against wavelength at full current (20~A, 26~A, and 23~A, respectively).
    }
    \label{fig:VECSELPowerMap}
\end{figure}

Figure~\ref{fig:VECSELPowerMap} summarizes the output power and wavelength tuning of the lasers: panels (a,b) show the 926~nm VECSEL, panels (c,d) the 910~nm VECSEL, and panels (e,f) the intracavity-doubled 455~nm VECSEL. All systems operate close to the diffraction limit, with $M^2<1.1$ for the NIR VECSELs and $M^2<1.04$ for the intracavity-doubled source. This beam quality enables efficient coupling ($>$85\%) into the subsequent enhancement cavities with comparatively simple mode-matching optics. The VECSELs used here were built in 2022--2023.

\subsection{Frequency noise, intensity noise, and tuning}
We characterize short-term frequency fluctuations with the side-of-fringe technique, using a confocal Fabry--P\'erot cavity with a free spectral range of 1.5~GHz and a finesse of 250. The laser is tuned to the slope of a cavity resonance, converting frequency fluctuations into amplitude variations of the transmitted signal. The cavity linewidth is approximately 6~MHz; at the half-maximum operating point, this corresponds to a frequency sensitivity of about 0.06~MHz per percentage point of the peak transmitted signal. The resulting single-sided frequency-noise power spectral density $S_\nu(f)$ is used to calculate the root-mean-square (RMS) frequency deviation,
\begin{equation}\label{eq:rmslinewidth}
   \nu_\mathrm{RMS} = \sqrt{ \int_{1/t}^{f_\mathrm{max}} S_\nu(f)\,df },
\end{equation}
where $t$ is the observation time and $f_\mathrm{max}$ is the measurement bandwidth set by the fast Fourier transform (FFT) analyzer.

\begin{figure*}[!htbp]
\centering
\includegraphics[width=\textwidth]{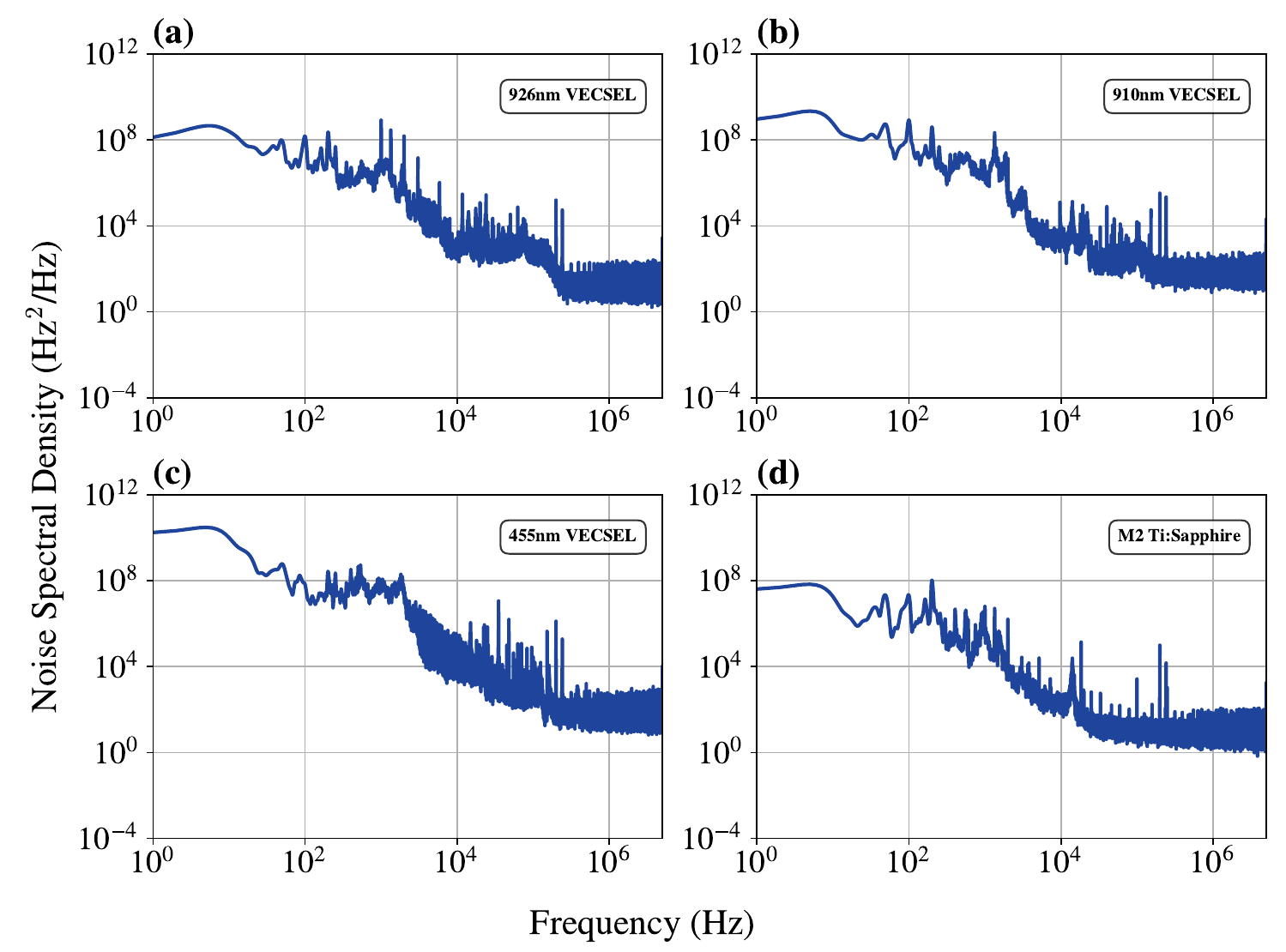}
\caption{Frequency-noise power spectral density of (a) the 926~nm VECSEL, (b) the 910~nm VECSEL, (c) the intracavity frequency-doubled 455~nm VECSEL, and (d) a Ti:sapphire reference laser (M~Squared SolsTiS), measured with a confocal Fabry--P\'erot cavity. In free-running operation, the VECSELs exhibit frequency noise close to that of the Ti:sapphire laser, showing that they provide the narrow linewidth required for laser cooling and spectroscopy without an external reference cavity. The corresponding RMS linewidths are summarized in Table~\ref{tab:linewidths}.}
\label{fig:NSD}
\end{figure*}

Because amplitude noise also contributes to the measured signal, this procedure gives an upper bound on the true frequency noise. The corresponding spectra are shown in Fig.~\ref{fig:NSD}(a)--(d) for the 926~nm VECSEL, the 910~nm VECSEL, the intracavity-doubled 455~nm VECSEL, and a Ti:sapphire reference laser, respectively. The frequency-noise data in Table~\ref{tab:linewidths} were recorded with the same 2023-built Vexlum Valo SF lasers.

\begin{table}[htbp]
  \centering
  \caption{Free-running RMS linewidths extracted from the frequency-noise spectra in Fig.~\ref{fig:NSD} using Eq.~(\ref{eq:rmslinewidth}).}
  \label{tab:linewidths}
  \begin{tabular}{@{}lcc@{}}
    \toprule
    \textbf{Laser system} & \multicolumn{2}{c}{\textbf{RMS linewidth, $\bm{\nu_\mathrm{RMS}}$}} \\
    \cmidrule(l){2-3}
     &$\bf{t=100~\upmu s}$ & $\bf{t=100~ms}$ \\
    \midrule
    926~nm VECSEL & 24~kHz & 197~kHz \\
    910~nm VECSEL & 20~kHz & 223~kHz \\
    455~nm intracavity-doubled VECSEL & 28~kHz & 386~kHz \\
    M~Squared SolsTiS Ti:sapphire & 9~kHz & 136~kHz \\
    \bottomrule
  \end{tabular}
\end{table}

\begin{figure*}
\centering
\includegraphics[width=\textwidth]{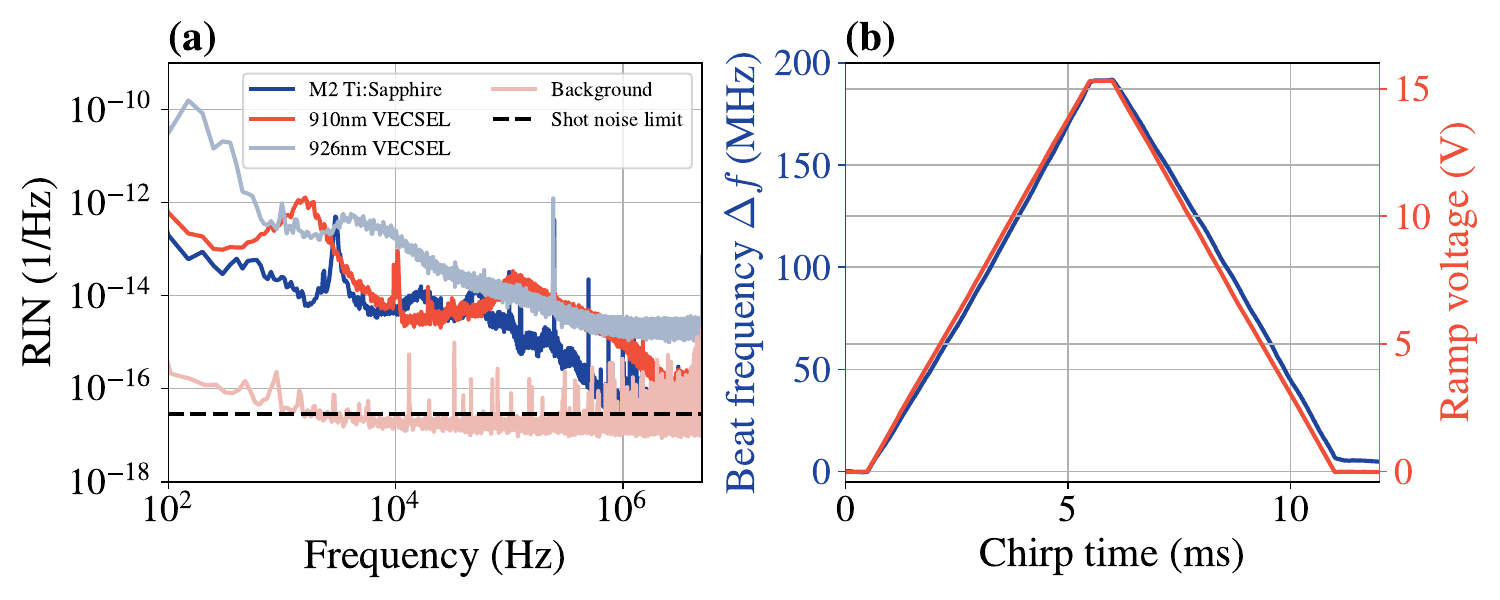}
\caption{(a) Relative intensity noise of the M~Squared SolsTiS, the 910~nm VECSEL, and the 926~nm VECSEL, measured with a biased detector with a bandwidth of 12~MHz. The detector background and the shot-noise limit are also shown. (b) Heterodyne measurement of a piezo-driven VECSEL frequency chirp. A 15~V voltage ramp produces a frequency excursion of approximately 200~MHz in 5~ms.}
\label{fig:IRRINNIR}
\end{figure*}

The relative intensity noise (RIN) is measured with a fast biased photodiode and compared with the Ti:sapphire laser in Fig.~\ref{fig:IRRINNIR}(a). The VECSELs also provide the tuning range required for spectroscopy and laser slowing. Coarse tuning is obtained with the birefringent filter, finer tuning over approximately 100~GHz is achieved by adjusting the etalon temperature, and continuous piezo tuning provides approximately 2~GHz of mode-hop-free scanning. The mode-hop-free range is not limited to the piezo stroke: applying a feed-forward correction to the etalon temperature that tracks the piezo ramp keeps the etalon transmission peak centered on the scanned cavity mode, and slow, mode-hop-free scans over tens of GHz are then achievable. The scan speed in this mode is set by the thermal response time of the etalon. Using a heterodyne measurement against the Ti:sapphire reference, we show that the VECSEL frequency can be chirped by 200~MHz in 5~ms [Fig.~\ref{fig:IRRINNIR}(b)], sufficient for frequency-chirped slowing of a cryogenic molecular beam of AlF~\cite{Padilla2025}. Even higher rates might be feasible but have not been explored in this work.

\section{Second-harmonic generation to blue wavelengths}\label{sec:blueshg}
NIR light from the 910--926~nm VECSELs is converted to the blue either by external resonant SHG in LBO or by intracavity SHG in the VECSEL cavity. The external LBO cavities provide the highest blue powers, reaching close to 4~W near 463~nm with cavity efficiencies up to 94\%, while the intracavity-doubled source requires fewer optical elements for mode matching to the DUV cavity and enables a more compact DUV system.

Throughout this work we distinguish three conversion efficiencies, since they answer different questions and are easily confused. The \emph{internal} efficiency is the ratio of the harmonic power generated inside the crystal to the fundamental power coupled into the cavity mode; it characterizes the nonlinear interaction alone. In the Brewster-cut BBO cavities, the uncoated output facet is oriented at Brewster's angle for the circulating fundamental. Because the generated DUV is orthogonally polarized, approximately 23\% is Fresnel reflected at this facet, giving $T_\mathrm{OC}=0.77$. The \emph{cavity} efficiency is the ratio of the harmonic power measured outside the cavity to the coupled-in fundamental power and is therefore lower than the internal efficiency by this extraction loss. The \emph{external} efficiency is the ratio of the harmonic power measured outside the cavity to the fundamental power incident on the cavity. Two effects set this fraction: the spatial mode matching, which is fixed by the input optics, and the impedance mismatch between the input-coupler transmission $T_0$ and the total round-trip loss. The latter is not constant. The nonlinear conversion itself acts as a loss that grows with circulating power, so a cavity chosen to be impedance matched at its design power is over-coupled at low power, and a fixed fraction of the incident light is then reflected rather than resonating. Unless stated otherwise, quoted efficiencies are cavity efficiencies; external efficiencies are reported alongside for comparison. The maximum values for the cavities characterized here are summarized in Table~\ref{tab:EfficiencyTable}.

\subsection{External LBO enhancement cavities}
External SHG is implemented with sealed bow-tie enhancement cavities (Agile Optic) containing 12-mm-long anti-reflection-coated LBO crystals. The cavity round-trip length is 406~mm and the waist in the nonlinear crystal is approximately 23~$\upmu$m ($1/e^2$ intensity radius).

The input-coupler transmission is $T_0 = 1.5\%$ and the cavity is stabilized using the H\"ansch--Couillaud locking scheme~\cite{Hansch1980}. Without SHG the finesse is approximately 300, consistent with residual round-trip losses $\delta\approx0.2\%$; during high-power conversion the nonlinear loss reduces the finesse to approximately 200.

\begin{figure*}[!htbp]
    \centering
    \includegraphics[width=\textwidth]{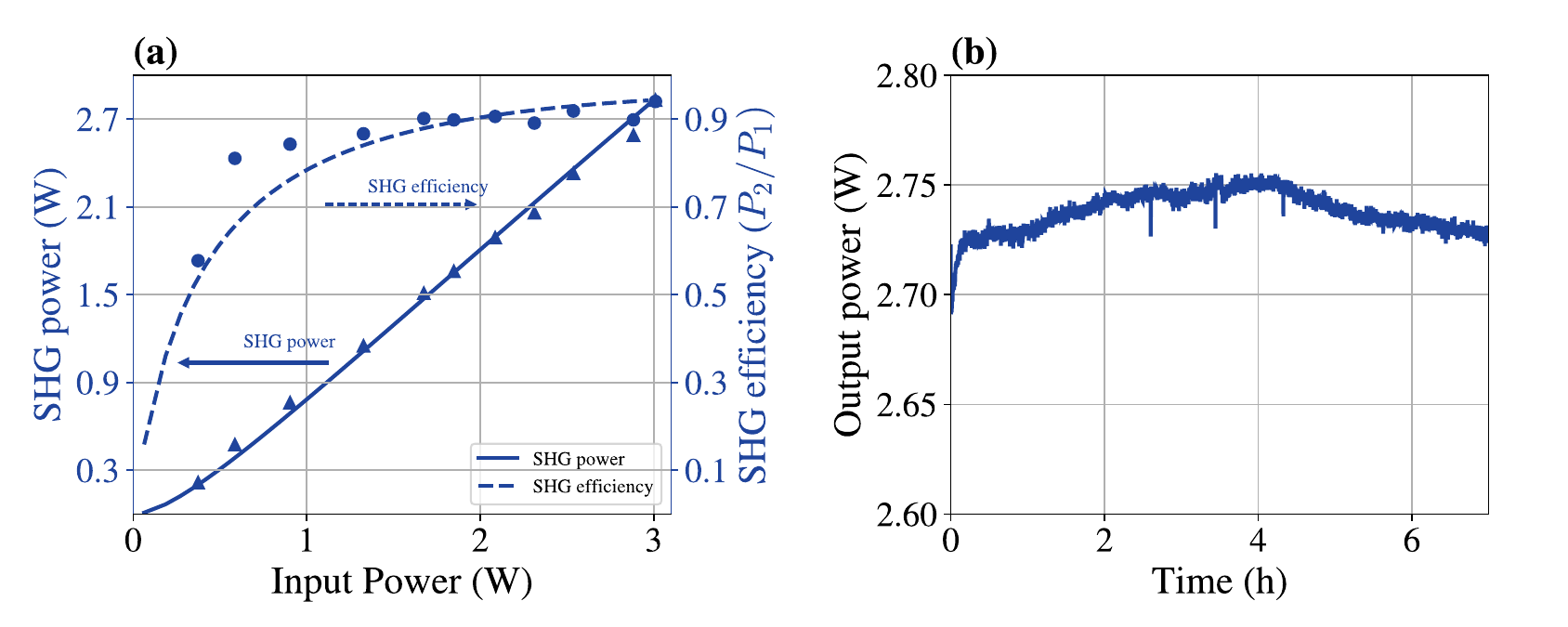}
    \caption{(a) SHG output power (triangles, solid curve; left axis) and conversion efficiency (circles, dashed curve; right axis) of the external LBO enhancement cavity for the 926~nm VECSEL system, as a function of the NIR power coupled into the cavity. The curves are fits to Eq.~(\ref{eq:shgfit}). The efficiency plotted is the cavity efficiency, $P_2/P_1$, which reaches 94\% at the highest measured input power. Because the LBO is AR coated at normal incidence there is no Brewster extraction loss, so the internal efficiency equals the cavity efficiency to within the residual AR reflection; referring the blue power to that incident on the cavity gives a maximum external efficiency of 88\%. (b) Output power of the locked cavity over seven hours under the reduced-power operating conditions used for the stability measurement.
    }
    \label{fig:BlueEfficiency}
\end{figure*}

Figure~\ref{fig:BlueEfficiency}(a) shows the generated blue power as a function of in-coupled NIR power for the 926~nm system. The data are fitted using the standard singly resonant SHG enhancement-cavity model, including power-dependent nonlinear conversion loss~\cite{Polzik1991,Jurdik2002}:

\begin{equation}\label{eq:shgfit}
    P_1=\frac{1}{4T_0\gamma^{1/2}}\left(\frac{P_2}{T_\mathrm{OC}}\right)^{1/2}\left[T_0+\delta+\left(\frac{P_2\gamma}{T_\mathrm{OC}}\right)^{1/2}\right]^2,
\end{equation}
where $P_1$ is the NIR power coupled into the cavity, $P_2$ the measured second-harmonic power outside the cavity, $T_0$ the input-coupler transmission, $T_\mathrm{OC}$ the total transmission from the generated harmonic inside the crystal to the measured output (including the Fresnel transmission of a Brewster-cut crystal), $\delta$ the residual linear round-trip loss, and $\gamma$ the single-pass SHG conversion coefficient. Thus, $P_2/T_\mathrm{OC}=\gamma P_\mathrm{c}^2$, where $P_\mathrm{c}$ is the circulating fundamental power.

The fit gives $\gamma=(4.94\pm0.41)\times10^{-5}~\mathrm{W}^{-1}$. The quoted uncertainty is statistical and does not include the systematic uncertainty in the input-coupler transmission $T_0$. The external LBO cavities produce up to 1.4~W at 455~nm from the 910~nm VECSEL and more than 2.7~W at 463~nm from the 926~nm VECSEL under the reduced-power operating conditions used for the stability measurement in Fig.~\ref{fig:BlueEfficiency}(b). The reduced power was chosen to limit the thermal load on the gain chip and preserve long-term device lifetime. When the 926~nm VECSEL is operated closer to its maximum output power of 5.4~W, the same external SHG stage produces close to 4~W at 463~nm. Across these operating conditions, the cavity efficiency peaks at 94\% of the coupled NIR power, while the external efficiency reaches 88\%. The cavity remains locked for seven hours with minimal drift.

\subsection{Comparison with intracavity blue generation}
\begin{figure*}
    \centering
    \includegraphics[width=\textwidth]{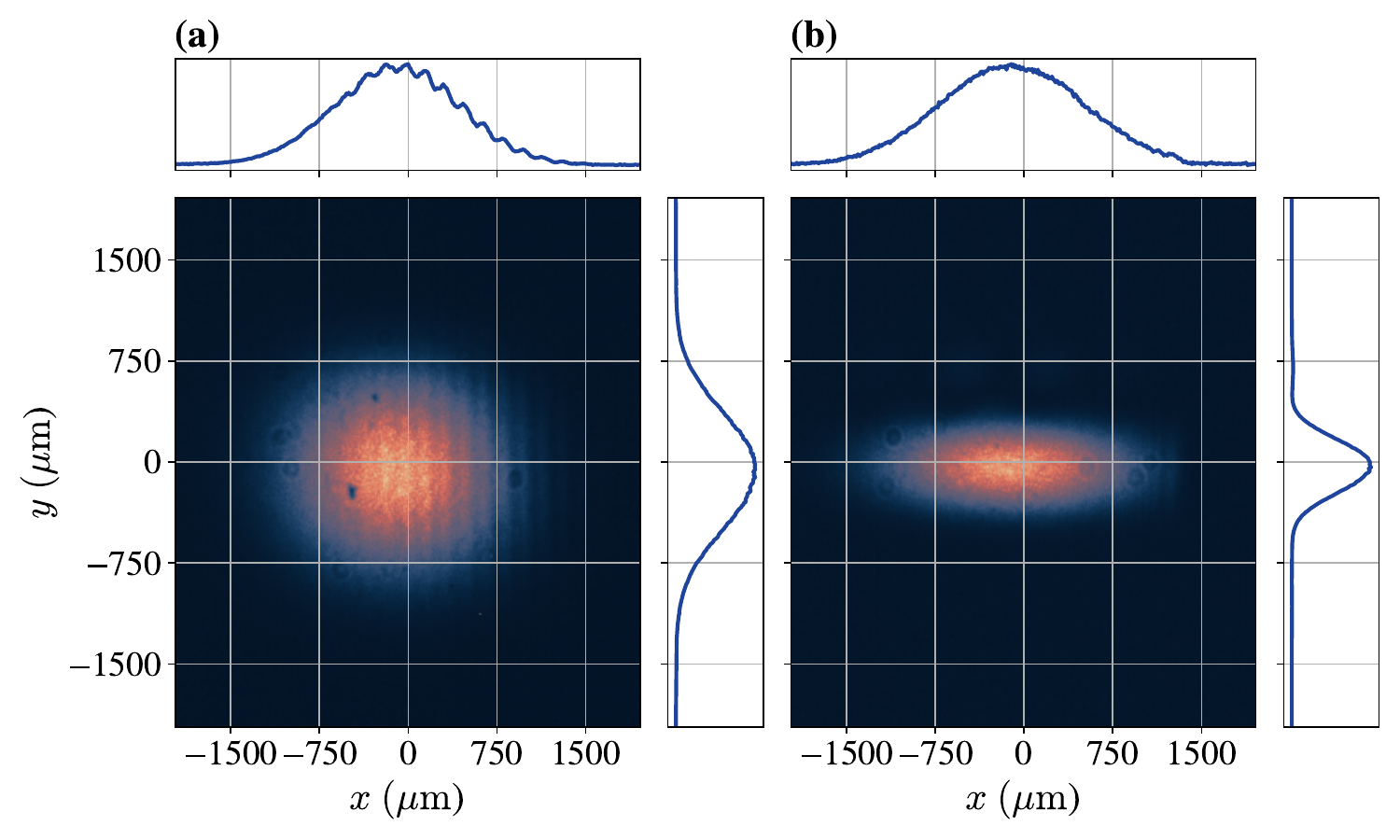}
    \caption{Beam profiles of the 455~nm light measured at the outputs of (a) the intracavity-doubled VECSEL and (b) the external LBO enhancement cavity, with profiles integrated along each axis. The intracavity-doubled source yields a near-circular mode (interference fringes from the beam profiler and neutral-density filters can be seen), whereas walk-off in the LBO makes the external-cavity output elliptical, requiring correction of the aspect ratio before the DUV stage.}
    \label{fig:BlueProfile}
\end{figure*}

The external LBO cavities provide higher blue output power, but their tightly focused type-I interaction produces an elliptical beam owing to spatial walk-off in LBO. As shown in Fig.~\ref{fig:BlueProfile}, the intracavity-doubled source produces a nearly circular 455~nm beam [Fig.~\ref{fig:BlueProfile}(a)], whereas the external LBO cavity produces a more elliptical output mode [Fig.~\ref{fig:BlueProfile}(b)]; the latter therefore requires additional cylindrical optics for efficient coupling into the DUV cavity. In the intracavity-doubled VECSEL, the high intracavity power permits weaker focusing in the LBO, and the VECSEL resonator filters higher-order spatial distortions. The main practical advantage of intracavity doubling is that it reduces the number of optical elements needed to mode match the blue beam to the DUV cavity and removes one actively locked enhancement cavity from the optical chain, reducing the size and complexity of the complete system.

\begin{figure}[ht]
    \centering
    \includegraphics[width=0.6\linewidth]{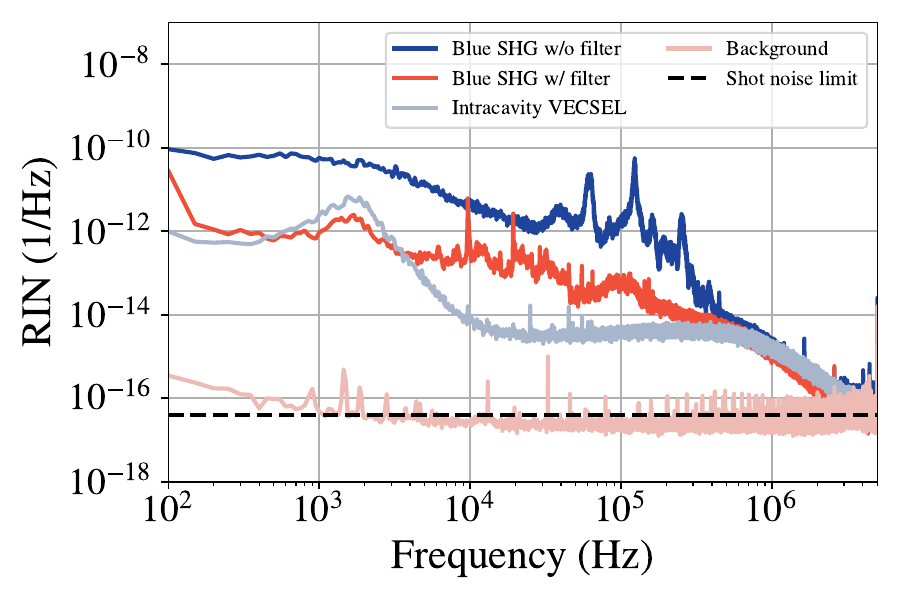}
    \caption{Relative intensity noise (RIN) of the two blue-generation schemes. Curves show the external LBO enhancement cavity before and after filtering, the intracavity-doubled VECSEL, and the detector background; the dashed line marks the shot-noise limit. The filtering (a twin-T notch in the error signal plus low-pass filtering of the piezo control voltage) suppresses the 60~kHz feature of the digital PID lock.
    }
    \label{fig:BlueRIN}
\end{figure}

The relative intensity noise of the two blue-generation schemes is compared in Fig.~\ref{fig:BlueRIN}. For the externally doubled system, a 60~kHz electronic feature associated with the digital proportional--integral--derivative (PID) controller is suppressed by adding a twin-T notch filter in the error-signal path and a 10~k$\Omega$ series resistor for low-pass filtering of the piezo drive. With this filtering, the RIN of the external SHG system approaches that of the intracavity-doubled VECSEL at frequencies above $\sim$100~kHz, though it remains somewhat higher in the 3--300~kHz band.

\section{Second-harmonic generation to the deep ultraviolet}\label{sec:duv}
\subsection{Focusing model and walk-off}
Efficient CW generation near 230~nm requires resonant second-harmonic generation of blue light in BBO. This wavelength range is substantially more challenging than the mature 266~nm technology based on the fourth harmonic of 1064~nm. The key reason is the loss of CLBO as the preferred deep-UV nonlinear crystal: CLBO, the established workhorse for high-power 266~nm generation, cannot be birefringently phase-matched for second-harmonic generation below about 237~nm~\cite{Nikogosyan2005}. Above this limit, CLBO is generally superior to BBO: it has smaller spatial walk-off, larger angular acceptance, higher effective conversion efficiency, and comparatively low two-photon absorption. These properties have enabled robust multi-watt CW generation at 266~nm~\cite{Sakuma2004} and 261.5~nm~\cite{Shaw2021} and watt-level operation near 240~nm~\cite{Burkley2019}. Below this SHG limit, BBO is still the only practical crystal for direct SHG, since type-I phase matching remains possible down to approximately 205~nm; fluoroberyllate (KBBF/RBBF) and fluorooxoborate crystals (NH$_4$B$_4$O$_6$F~\cite{Zhang2026VUV}, CsB$_4$O$_6$F) phase-match to still shorter wavelengths, but remain difficult to grow and are not commercially available. CLBO itself remains available below 237~nm through sum-frequency mixing rather than doubling, which is how CW light at 193~nm is produced~\cite{Sakuma2011}; that route trades the crystal problem for a second single-frequency laser and an extra enhancement cavity, and we do not pursue it here. Although BBO can also support watt-level CW generation at 266~nm~\cite{Oka1995}, operating it close to its short-wavelength limit introduces severe penalties, which have so far kept practical CW powers below 230~nm in the sub-watt regime.

These penalties arise from several coupled effects. Near 230~nm, BBO exhibits large birefringent walk-off, with $\rho L\simeq0.75$~mm for a $L=10$~mm crystal, which limits the useful interaction length and distorts the generated DUV beam. Here $\rho$ is the walk-off angle. At the same time, the angular and spectral acceptances become small, the Boyd--Kleinman focusing factor is reduced, and DUV absorption rises rapidly. The larger photon energy, about 5.4~eV at 230~nm compared with 4.7~eV at 266~nm, further accelerates heating, color-center formation, and DUV-induced surface contamination~\cite{Arnold2022UVContamination, hollenshead2019, hovisOpticalDamagePart1994, warburton1992, dubietis2000}. As a result, the single-pass conversion efficiency is intrinsically low and must be recovered with an enhancement cavity, but the achievable build-up is itself limited by the same absorption, thermal-lensing, and degradation mechanisms. The central design problem is therefore to maintain useful nonlinear conversion while reducing the local optical intensity that drives bulk damage, color-center formation, and thermal lensing. We address this by comparing spherical and elliptical focusing geometries in Brewster-cut BBO, before introducing a normal-incidence AR-coated BBO cavity.

For the BBO stage, the single-pass SHG power is written as~\cite{Boyd1968}
\begin{equation}\label{eq:shgeff}
    P_2 = K P_\mathrm{c}^2\, l\, k_1\, h(B,\xi_x,\xi_y)=\gamma P_\mathrm{c}^2,
\end{equation}
where $P_\mathrm{c}=T_0\mathcal{F}^2P_1/\pi^2$ is the fundamental circulating power in the absence of nonlinear conversion, expressed in terms of the cold-cavity finesse $\mathcal{F}$; for perfect impedance matching, this reduces to $P_\mathrm{c}/P_1\simeq\mathcal{F}/\pi$. Here, $l$ is the crystal length, $k_1$ is the fundamental wave number in the crystal, and $h(B,\xi_x,\xi_y)$ is the Boyd--Kleinman factor generalized to elliptical focusing~\cite{Steinbach1996,Freegarde1997}. The constant
\begin{equation}\label{eq:Kconstant}
    K = \frac{2\omega_1^2 d_\mathrm{eff}^2}{\pi\epsilon_0 c^3 n_1^2 n_2}
\end{equation}
contains the material parameters: $\omega_1$ is the fundamental angular frequency, $d_\mathrm{eff}$ the effective nonlinear coefficient, and $n_1$, $n_2$ the refractive indices at the fundamental and harmonic. The focusing parameters are $\xi_{x,y}=l/(w_{x,y}^2 k_1)$, where $w_{x,y}$ are the fundamental waist radii and $x$ denotes the walk-off direction. The walk-off parameter is $B=\rho\sqrt{lk_1}/2$ with
\begin{align}\label{eq:h}
    h(B,\xi_x,\xi_y) &= \frac{\sqrt{\xi_x\xi_y}}{l^2}\nonumber\\
    &\times \int_0^l \int_0^l \frac{e^{i\Delta k(z'-z)}e^{-4B^2\xi_x(z'-z)^2/l^2}\,dz\,dz'}
    {\sqrt{1+i\tau_x}\sqrt{1+i\tau_y}\sqrt{1-i\tau_x'}\sqrt{1-i\tau_y'}},
\end{align}
where $\Delta k=k_2-2k_1$ is the phase mismatch (optimized numerically for each geometry), $\tau_{x,y}=2(z-l/2)/(k_1 w_{x,y}^2)$, and primed variables are evaluated at $z'$; the focus is taken at the crystal center.

Elliptical focusing is used here to lower the local fundamental intensity in the BBO crystal while retaining comparable single-pass conversion efficiency in the large-walk-off regime~\cite{Steinbach1996,Gumm2025}. The waist is expanded along the walk-off axis and kept tight in the orthogonal direction, so that the nonlinear polarization is induced over a larger transverse area and the generated DUV field remains overlapped with the fundamental over a longer effective interaction length. All three DUV cavities use 10~mm-long BBO crystals cut for type-I phase matching. Throughout this section $P_\mathrm{c}$ denotes the circulating fundamental power, related to the generated harmonic by $P_2/T_\mathrm{OC}=\gamma P_\mathrm{c}^2$. At 460~nm the phase-matching angle is $\theta_\mathrm{pm}=60.4^\circ$, giving $\rho=75$~mrad ($\rho L=0.75$~mm), $B=18.0$, and an effective nonlinear coefficient $d_\mathrm{eff}=1.23$~pm/V, about 60\% of its value at 532~nm, because the $d_{22}\cos\theta_\mathrm{pm}$ term collapses at large phase-matching angle. The fundamental waists used in this work are approximately $w_x\times w_y = 262~\upmu\mathrm{m}\times68~\upmu\mathrm{m}$ for the elliptical Brewster-cut cavity, $44~\upmu\mathrm{m}\times67~\upmu\mathrm{m}$ for the spherical Brewster-cut cavity, and $60~\upmu\mathrm{m}\times60~\upmu\mathrm{m}$ for the AR-coated BBO cavity; these and the derived conversion coefficients are collected in Table~\ref{tab:gammadelta}. Geometric astigmatism is compensated through the incidence angles on curved mirrors M3 and M4.

For these waists, the elliptical geometry reduces the peak fundamental intensity sixfold at fixed circulating power. This does not result in a reduction of the DUV intensity at the output facet. The calculated full widths are $719~\upmu\mathrm{m}\times57~\upmu\mathrm{m}$ for the cavity with a spherical focus and $838~\upmu\mathrm{m}\times55~\upmu\mathrm{m}$ for the elliptical version. The spherical walk-off profile is nearly flat-topped (super-Gaussian order 11), whereas the elliptical profile remains near Gaussian (order 1.66). Second-moment propagation gives $(M_x^2,M_y^2)=(2.47,1.00)$ and $(1.09,1.00)$, respectively.

\begin{figure}[ht]
    \centering
    \includegraphics[width=0.75\linewidth]{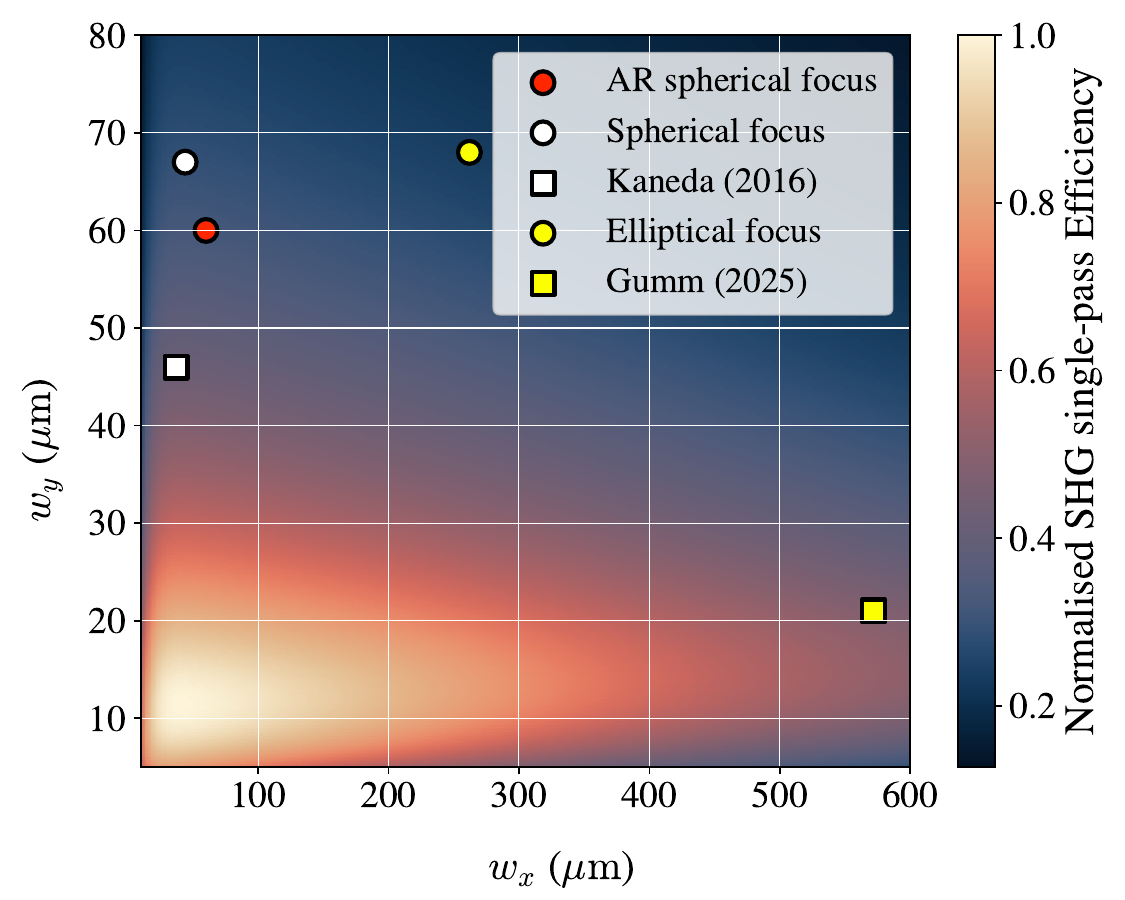}
    \caption{Calculated Boyd--Kleinman factor $h(B,\xi_x,\xi_y)$ for second-harmonic generation near 230~nm in a 10~mm BBO crystal, normalized to its maximum value, as a function of the fundamental (455--463~nm) waist radii $w_x$ (walk-off direction) and $w_y$ inside the crystal. Markers indicate the geometries studied in this work and reported in the literature~\cite{Kaneda2016,Gumm2025}; the corresponding values of $h$ are given in Table~\ref{tab:gammadelta}. In the large-walk-off regime, the waist along the walk-off axis can be relaxed substantially while retaining comparable conversion efficiency.
    }
    \label{fig:SHGeffElliptical}
\end{figure}

The important intensity reduction occurs inside the crystal: the DUV is generated at a lower local source density, reducing the integrated bulk optical dose that drives photo-induced degradation, heating, and thermal lensing. The distributed nonlinear source also improves the generated DUV mode profile, which is an important practical advantage of elliptical focusing near 230~nm. At these wavelengths, tightly focused spherical BBO cavities are particularly sensitive to walk-off-induced phase mismatch, local defects, and thermal gradients, all of which can distort both the circulating fundamental mode and the generated DUV beam. By spreading the nonlinear interaction along the walk-off axis, the elliptical geometry reduces this sensitivity and produces a cleaner DUV mode that is easier to deliver and reshape. This matters for applications that require high interferometric contrast, tight focusing, stable beam delivery, or uniform illumination. These trade-offs are reflected in the calculated Boyd--Kleinman factor shown in Fig.~\ref{fig:SHGeffElliptical}: the waist along the walk-off direction can be enlarged substantially while retaining comparable conversion efficiency.

\FloatBarrier
\subsection{Fixing the single-pass conversion coefficient}\label{sec:gammafix}

Rather than treating the single-pass coefficient $\gamma$ as a free parameter, we constrain it to its Boyd--Kleinman value, computed from Eqs.~(\ref{eq:shgeff})--(\ref{eq:h}) for the measured waists, and let the residual round-trip loss $\delta$ carry a small cavity-to-cavity variation. In order to compare our cavities directly with the closest published geometry, we normalize $\gamma$ to the calculated value in Ref.~\cite{Kaneda2016} rather than to a tabulated $d_\mathrm{eff}$, which is known here only to about 20\%. Kaneda et al. calculate $\gamma=0.62\times10^{-4}~\mathrm{W}^{-1}$ for a 10~mm crystal and a cavity waist of $w_x\times w_y=37~\upmu\mathrm{m}\times46~\upmu$m; this was not an independent single-pass measurement. Scaling their value by the ratio of $h(B,\xi_x,\xi_y)$ between their geometry and ours gives $4.7\times10^{-5}~\mathrm{W}^{-1}$ for our AR-coated cavity. The material constants cancel in this ratio, so the transported value depends only on the focusing geometry, and it agrees with the value obtained from our own data. Table~\ref{tab:gammadelta} lists the Boyd--Kleinman factor $h$ and the resulting $\gamma$ for each geometry, together with the calculated reference value from Ref.~\cite{Kaneda2016}.

The loss is independently bounded by the measured cold-cavity finesse, $\mathcal{F}$. In the absence of conversion, the DUV cavities typically have $\mathcal{F}=250$--$300$, corresponding to $T_0+\delta=2.1$--$2.5\%$. The nominal input-coupler transmission is $T_0=1.5(2)\%$. With $\gamma$ constrained, reproducing the measured harmonic powers requires $\delta=0.79\%$, $0.71\%$, and $0.66\%$ for the spherical Brewster-cut, elliptical Brewster-cut, and AR-coated cavities, respectively. These values imply finesses of 271, 281, and 288, consistent with the measured range within the uncertainty in $T_0$.

\begin{table}[!htb]
  \centering
  \caption{Single-pass conversion coefficient $\gamma$ constrained to its Boyd--Kleinman value and residual round-trip loss $\delta$ required to reproduce the scanned harmonic power, together with the cold finesse $\mathcal{F}$ implied by $T_0=1.5\%$ and the circulating power $P_\mathrm{c}$ at the maximum scanned output. Waists are quoted as walk-off axis $\times$ orthogonal axis. All crystals are 10~mm long.}
  \label{tab:gammadelta}
  \begin{tabular}{@{}lcccccc@{}}
    \toprule
    \textbf{DUV cavity} & $\bm{w_x\times w_y}$ & $\bm{h}$ & $\bm{\gamma}$ & $\bm{\delta}$ & $\bm{\mathcal{F}}$ & $\bm{P_\mathrm{c}}$ \\
     & ($\upmu$m) & ($10^{-2}$) & ($10^{-5}~\mathrm{W}^{-1}$) & (\%) & & (W) \\
    \midrule
    Brewster-cut, spherical  & $44\times67$  & 1.46 & 4.27 & 0.79 & 271 & 146 \\
    Brewster-cut, elliptical & $262\times68$ & 1.21 & 3.52 & 0.71 & 281 & 173 \\
    AR-coated, spherical     & $60\times60$  & 1.62 & 4.73 & 0.66 & 288 & 145 \\
    \addlinespace
    Ref.~\cite{Kaneda2016}   & $37\times46$  & 2.13 & 6.20 & --   & --  & --  \\
    \bottomrule
  \end{tabular}
\end{table}

\FloatBarrier
\subsection{Damage mechanisms and long-term mitigation}

BBO degradation during CW operation near 230~nm is driven by coupled photochemical, electronic, and photothermal processes~\cite{WuTPA2001,kondratyuk2002,dubietis2000,turcicova_laser_2022}. The crystal is exposed for many hours to high circulating blue intensity and generated DUV power, with the most severe conditions near the output facet. Relevant mechanisms include DUV-induced cracking of adsorbed hydrocarbons, linear and defect-assisted absorption of the circulating blue and DUV fields, color-center formation, and heat deposition leading to thermal lensing and stress. Nonlinear absorption may also contribute but is expected to be small compared with linear absorption in our operating intensity regime. For Brewster-cut crystals, the output facet is particularly vulnerable because it is an uncoated BBO surface where the DUV power is largest and Fresnel-reflected DUV light further increases the local optical dose. We note, however, that whereas carbonaceous films are well established on DUV mirrors and windows, they have not yet been chemically identified on a BBO facet; hydrocarbon photochemistry on the crystal surface therefore remains a plausible but unconfirmed degradation channel.

Environmental control is therefore as important as the optical design. The cavity housing forms a sealed, clean enclosure with a continuous filtered dry-air purge to reduce volatile organic compounds, water vapor, and particulates. This is critical because 230~nm photons efficiently crack adsorbed hydrocarbons. At atmospheric pressure, even a 1~ppb hydrocarbon background corresponds to an impingement flux of order $10^{14}$--$10^{15}~\mathrm{cm^{-2}s^{-1}}$, so absorbing carbonaceous layers can form on experimentally relevant timescales even when the sticking and photochemical conversion probabilities are far below unity. Heating the BBO reduces the residence time and sticking probability of weakly bound adsorbates, favoring desorption before photochemical conversion can occur~\cite{hovisOpticalDamagePart1994,kunz_experimentation_2000, cooper_cavity-enhanced_2018, zhao_high-power_2017}. Cavity materials, cleaning procedures, purge purity, flow geometry, and crystal temperature must therefore all be treated as part of the DUV laser design.

Dry air also provides an \emph{in situ} cleaning pathway. Below about 242~nm, DUV photodissociation of molecular oxygen produces ozone and atomic oxygen~\cite{AMORUSO1996}. At 227.5~nm, the measured O$_2$ cross-section implies that each watt of DUV produces about $4\times10^{13}$ oxygen atoms per second per centimeter of beam path; with our 1~L/min purge, the estimated steady-state ozone density near the output facet is of order $10^{13}~\mathrm{cm^{-3}}$. These reactive species oxidize hydrocarbon fragments into volatile products, whereas pure nitrogen removes this cleaning pathway. A clean, very dry, continuously refreshed oxygen-containing purge therefore balances contaminant removal against surface attack.

Elliptical focusing complements environmental control by lowering the peak circulating blue intensity throughout the crystal while preserving useful SHG efficiency~\cite{Kiefer2019, Preiler2019, Gumm2025}. In our operating regime, the main expected benefit is lower local heat deposition from linear and defect-assisted absorption and smaller thermal gradients; the exit-facet DUV intensity changes only modestly, as discussed above.

Crystal quality, surface preparation, and temperature further determine long-term robustness. We have obtained good performance with commercially available BBO crystals, provided that the cavity is clean, dry, and well aligned and the crystal is heated. Czochralski (CZ)-grown BBO is therefore not an absolute requirement for successful operation, but it improves stability: in our experience it shows weaker thermal lensing, smaller power-dependent shifts of the optimum cavity mode matching, and more reproducible high-power operation than typical top-seeded solution-growth material. 
Operation above 100$^\circ$C further improves stability by reducing adsorption of water and hydrocarbons on the mildly hygroscopic surface, limiting hydroxylation and carbonaceous film growth; it may also accelerate defect relaxation or charge recombination, suppressing slow color-center or trapped-charge accumulation~\cite{takachiho_ultraviolet_2014,Takahashi2010}. In closely related CW second-harmonic generation at 257~nm, a tightly focused spherical cavity (32~$\upmu$m waist) degraded within minutes, and heating alone suppressed this degradation only at low power, whereas an elliptical focus combined with operation above 100$^\circ$C enabled watt-level output for hours without degradation~\cite{Kiefer2019,Gumm2025}. Because the focusing, temperature, and other operating conditions changed together, this comparison supports the combined mitigation strategy but does not isolate elliptical focusing as the cause of the improvement.

Surface preparation is especially important because mechanical polishing can leave a chemically modified near-surface layer containing subsurface damage, residues, hydroxylated material, and adsorbed carbon. Ion-beam etching removes this altered layer and provides a cleaner interface for either bare Brewster operation or subsequent coating. For normal-incidence crystals, AR coatings are deposited after ion-beam etching. These coatings are promising because they can both reduce Fresnel loss and protect the BBO surface from moisture and hydrocarbon adsorption. Their full long-term performance under high-power CW DUV exposure is still being evaluated.

The broader UVQuanT record supports the robustness of this approach: over four years, 14 DUV cavities in 12 laser systems at six European laboratories have operated with minimal servicing, using BBO crystals from multiple suppliers, growth methods, and surface-preparation routes. As one example, the elliptical Brewster-cut cavity has routinely produced 400--600~mW of DUV light over nearly four years; its crystal changes were made mainly to compare suppliers and preparation procedures rather than because replacement was required. This experience shows that high-power CW generation near 230~nm is feasible when reduced local intensity, elevated-temperature operation, stringent environmental cleanliness, careful surface preparation, and continuous dry-air purging are combined. CZ-grown material and ion-beam-prepared or passivated surfaces further improve the operational margin, especially at the highest powers, but are not the only route to robust DUV generation.

A complementary mitigation, which we note as an outlook, is to slowly translate or dither the crystal rather than accumulating dose at a single spot. This is a promising route to further extend crystal lifetime.

\subsection{Brewster-cut BBO cavities}\label{sec:brewster}
\begin{figure}
    \centering
    \includegraphics[width=\columnwidth]{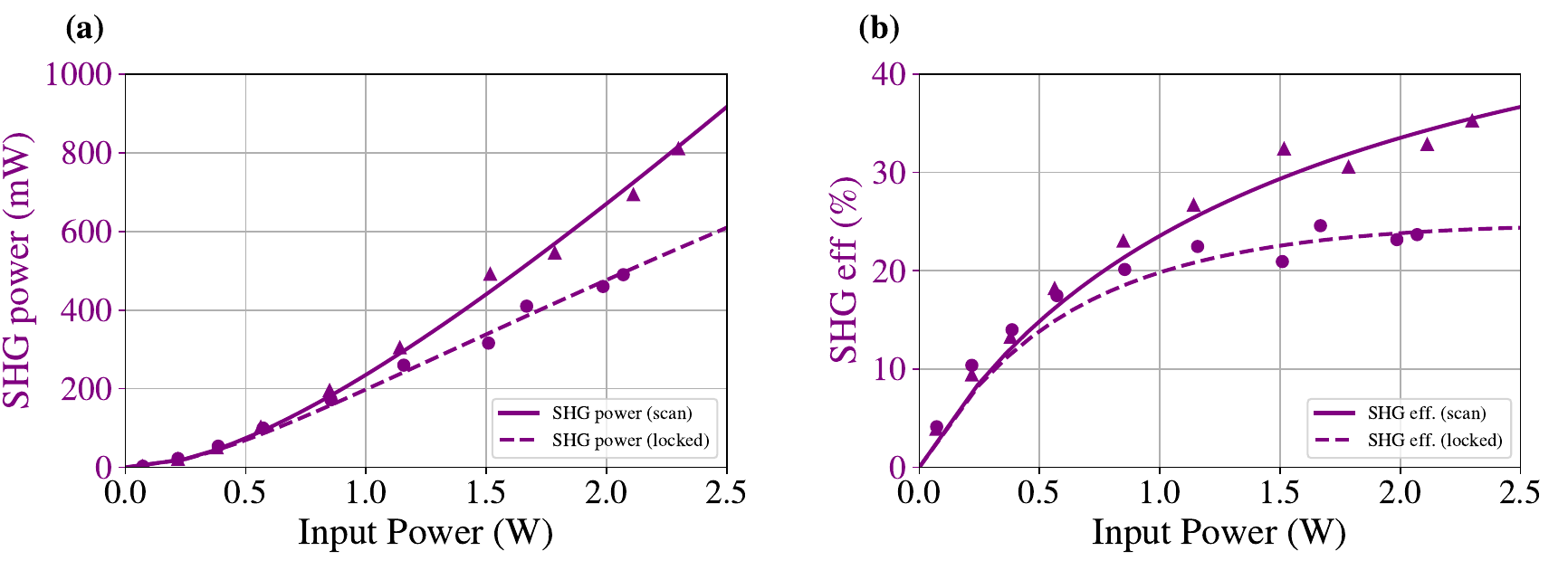}

          \caption{DUV performance of the near-230~nm elliptical-focus Brewster-cut BBO enhancement cavity with the cavity length scanned (triangles, solid curves) and locked (circles, dashed curves). (a) DUV output power as a function of in-coupled blue power. The solid curve is Eq.~(\ref{eq:shgfit}) with $\gamma$ constrained to its Boyd--Kleinman value of $3.52\times10^{-5}~\mathrm{W}^{-1}$ and a residual round-trip loss $\delta=0.71\%$ (Table~\ref{tab:gammadelta}), consistent with the measured cold finesse. The dashed curve uses the same $\gamma$ and $\delta$ with the power-dependent term $\alpha P_2/T_\mathrm{OC}$ added, $\alpha=5.4\times10^{-3}~\mathrm{W}^{-1}$, corresponding to an effective round-trip loss of $1.6\%$ at the highest in-coupled power. We attribute this additional loss to a thermally induced change in mode overlap in the more alignment-sensitive elliptically focused cavity rather than to absorption of the generated DUV light. (b) Corresponding cavity SHG efficiency in scanned and locked operation, reaching 35\% and 24\%, respectively, at the highest in-coupled blue power. Adding back the 23\% Fresnel loss at the Brewster output facet gives internal efficiencies of 46\% and 31\%. Referring the delivered DUV to the blue power incident on the cavity gives external efficiencies of 30\% and 18\%. The coupled-in power is determined from the reflected power, so the impedance mismatch appears only in the external values.}
\label{fig:V3UVEfficiency}
\end{figure}

Figure~\ref{fig:V3UVEfficiency} shows the DUV output power [panel~(a)] and conversion efficiency [panel~(b)] of the elliptical-focus cavity as a function of in-coupled blue power for scanned and locked operation.
With $\gamma$ constrained to the Boyd--Kleinman value of Table~\ref{tab:gammadelta}, the scanned data are described by Eq.~(\ref{eq:shgfit}) with the residual round-trip loss $\delta=0.71\%$, consistent with the measured cold finesse. In locked operation the same $\gamma$ and $\delta$ reproduce the data once the power-dependent term $\alpha P_2/T_\mathrm{OC}$ is added to the square bracket, with $\alpha=5.4\times10^{-3}~\mathrm{W}^{-1}$; the effective round-trip loss then reaches $1.6\%$ at the highest in-coupled power, of which $0.71\%$ is passive, $0.47\%$ is conversion, and $0.38\%$ is the excess described by $\alpha$. This reduction is mitigated by heating the crystal above 100$^\circ$C and by optimizing the separation of the curved mirrors that focus the fundamental into the crystal, consistent with a thermally induced change in effective mode overlap rather than loss of generated DUV power. To reach the highest powers, the cavity, output coupler, and phase-matching angle are re-optimized while locked; this recovers much of the difference from the scanned curve and yields the maximum powers reported below.

In our Brewster-cut BBO crystals, the Fresnel loss of the DUV light is about 23\%. Accordingly, in Eq.~(\ref{eq:shgfit}), we use $T_\mathrm{OC}=0.77$ for the Brewster-cut cavities. This is the loss that separates the internal and cavity efficiencies for this geometry, and it is why an internal efficiency quoted for a Brewster-cut crystal overstates the power that reaches the experiment. Taking this loss together with the coupling of the incident light into the resonant mode, the external blue-to-DUV efficiency of the elliptical cavity is 18\% locked and 30\% scanned, against cavity efficiencies of 24\% and 35\% in Fig.~\ref{fig:V3UVEfficiency}(b) and internal efficiencies of 31\% and 46\%. The coupled-in power $P_1$ is obtained from the incident and reflected powers.

\begin{figure*}
    \centering
    \includegraphics[width=\textwidth]{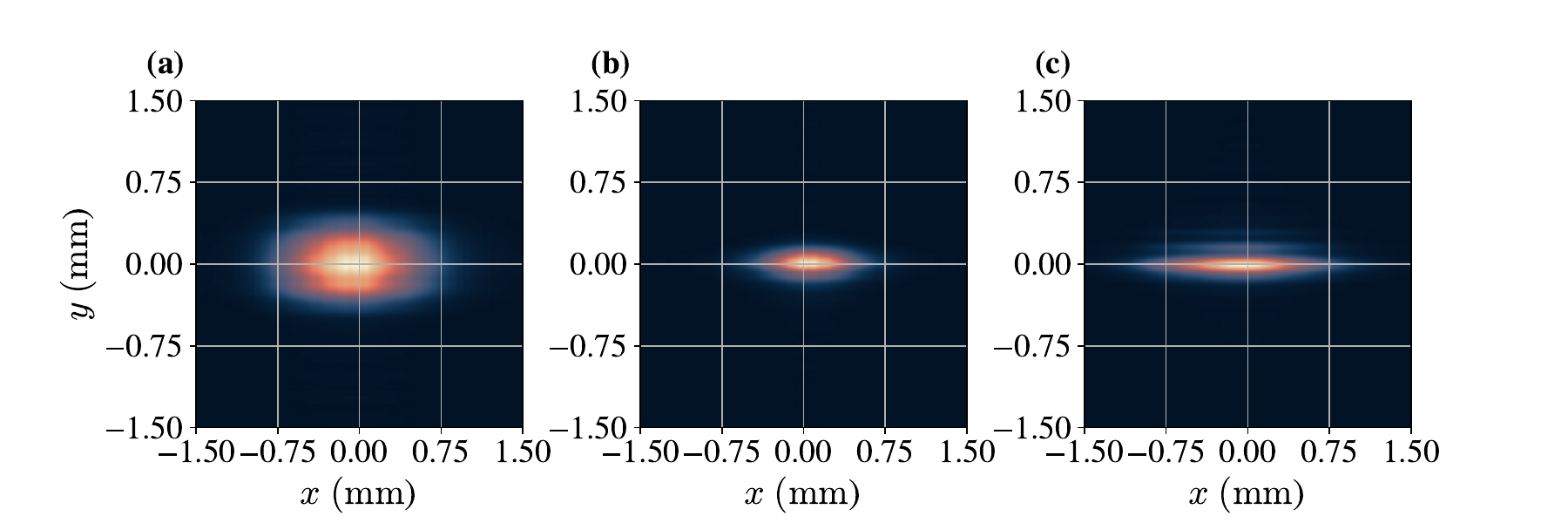}
    \caption{DUV beam profiles near 230~nm, measured 28~cm after the output window, for (a) the elliptical-focus Brewster-cut cavity, (b) the normal-incidence AR-coated cavity, and (c) the spherical-focus Brewster-cut cavity.}
    \label{fig:UVprofile}
\end{figure*}

An advantage of the elliptical geometry is the quality of its DUV beam. Figure~\ref{fig:UVprofile} shows the beam profiles measured 28~cm after the output window for the elliptical Brewster-cut cavity [panel~(a)], the normal-incidence AR-coated cavity (see the following subsection) [panel~(b)], and the spherical Brewster-cut cavity [panel~(c)]. The different transverse sizes at the same observation distance reflect differences in the generated source dimensions, divergence, walk-off distribution, and output geometry of the three cavities. The tightly focused spherical Brewster-cut cavity delivers a fragmented mode, whereas relaxing the focus along the walk-off axis produces a smooth, well-defined spot; the normal-incidence AR-coated cavity likewise delivers a clean, compact mode. A clean beam improves delivery for applications requiring tight focusing, interferometric stability, or uniform illumination, while the relaxed focus also lowers the local intensity in the BBO and mitigates bulk degradation and thermal lensing.

The spherical-focus Brewster-cut cavity behaves differently. We measure output powers up to 700~mW of DUV light for 2.6~W of blue power incident on the cavity, of which 75--80\% is coupled into the resonant mode, giving a cavity efficiency of 34\%, an external efficiency of 27\%, and an internal efficiency of 44\%. The long-term measurements presented in Section~\ref{sec:longterm} were recorded at slightly lower operating powers and include a continuous 70~h run. After the phase matching was re-optimized with the cavity locked at maximum input power, the scanned and locked values agree within the measurement uncertainty. Smaller differences remain at other input powers, but they are less pronounced than for the elliptical cavity; estimates of the thermal-lens focal length indicate that the in-coupling waist of the spherical cavity is less sensitive to the thermally induced lens.

Overall, the two Brewster-cut geometries present a trade-off, though not the one the scanned data alone would suggest. Under scanned operation, the elliptical cavity is marginally the more efficient of the two, 35\% against 34\%; when locked, the ordering reverses, 24\% against 34\%. This difference likely includes power-dependent impedance matching as well as thermally induced changes in effective mode overlap; the empirical $\alpha$ term describes the observed reduction but does not identify a unique loss channel. The elliptical cavity typically shows small higher-order spatial-mode contributions. Small alignment errors in the cylindrical mirrors, or finite waist and phase-matching offsets, reduce the effective nonlinear overlap without strongly changing the measured cavity finesse; the cavity is correspondingly harder to align. In return, it provides a cleaner delivered DUV mode and reduces the local fundamental intensity. The sustained spherical-cavity measurements show, however, that a tighter focus does not by itself imply degradation.

\subsection{Normal-incidence AR-coated CZ-grown BBO cavity}\label{sec:arcavity}
The Brewster-cut BBO cavities are robust, but the geometry limits the extractable DUV power: for our crystal cut and orientation, approximately 23\% of the generated DUV light is Fresnel-reflected at the output facet rather than delivered to the experiment. We therefore implemented a normal-incidence BBO cavity in which both crystal facets are AR coated for the fundamental and second-harmonic wavelengths, and the DUV is extracted with an intracavity dichroic output coupler.

This approach was motivated in part by recent high-power results with elliptically focused BBO enhancement cavities at longer UV wavelengths~\cite{Gumm2025}. Operation near 230~nm is more demanding because the higher photon energy increases both bulk and surface degradation rates. We therefore use CZ-grown BBO selected for low DUV absorption and high optical homogeneity. Before coating, the facets were ion-beam etched using the surface-preparation strategy discussed above, providing a cleaner interface for the dielectric stack. The coating was designed for low residual reflection at both wavelengths for high-intensity CW DUV operation.

\begin{figure}
    \centering
    \includegraphics[width=\columnwidth]{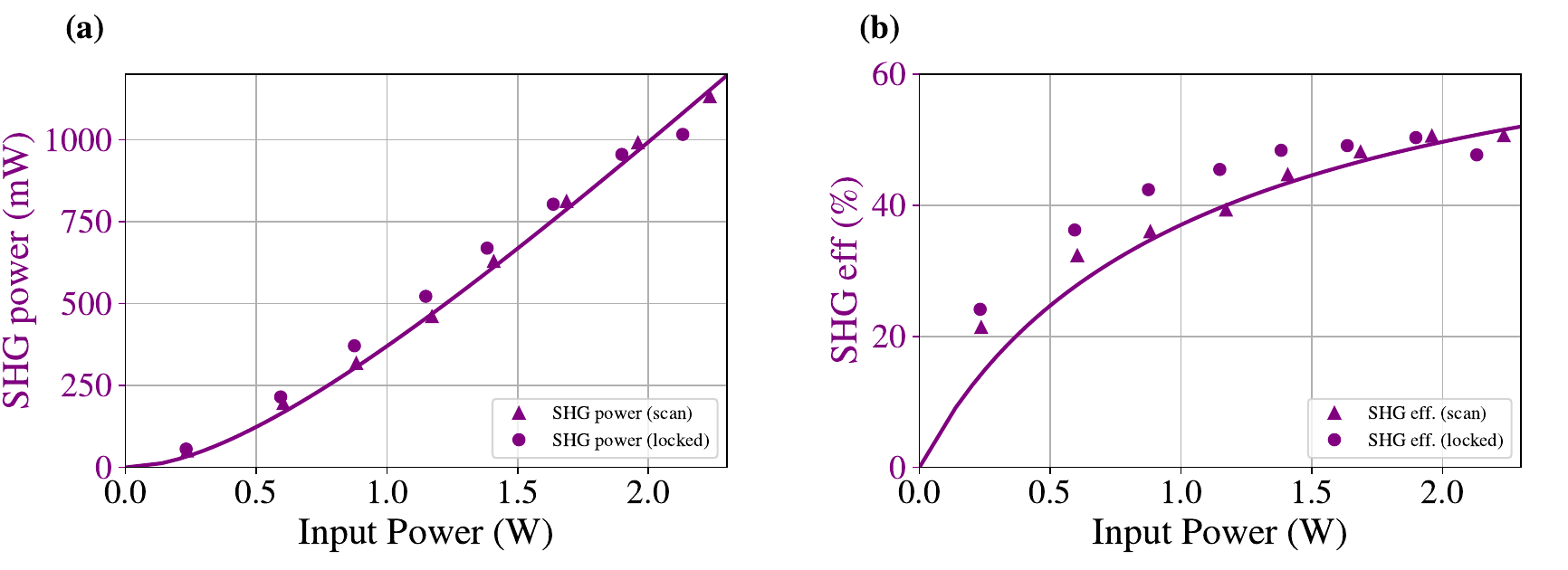}
    \caption{DUV performance of the near-230~nm spherical-focus, normal-incidence AR-coated BBO enhancement cavity. (a) DUV output power versus in-coupled blue power. (b) Corresponding cavity SHG efficiency. The solid curve is Eq.~(\ref{eq:shgfit}) with $\gamma$ constrained to its Boyd--Kleinman value of $4.73\times10^{-5}~\mathrm{W}^{-1}$ and $\delta=0.66\%$ (Table~\ref{tab:gammadelta}), consistent with the measured cold finesse. No power-dependent correction is required; after alignment and phase matching are optimized at the relevant operating power, the locked value agrees with the scan there within the measurement uncertainty. The cavity efficiency reaches 51\% with the cavity locked. Both facets are AR coated at normal incidence, so there is no Brewster extraction loss and the internal efficiency equals the cavity efficiency to within the residual AR reflection; the external efficiency is 44\%. The three conventions are defined in Section~\ref{sec:blueshg}.}
    \label{fig:V3UVEfficiency_AR}
\end{figure}

The normal-incidence geometry removes the Brewster-facet extraction loss while retaining the large-focus strategy used for the uncoated cavities. Figure~\ref{fig:V3UVEfficiency_AR} shows its DUV output power [panel~(a)] and conversion efficiency [panel~(b)] as a function of in-coupled blue power. After alignment and phase matching were optimized for the relevant operating power, the locked values agree with the scan within the measurement uncertainty at that point; this does not imply that a single fixed alignment makes the two datasets identical over the full power range. 

No additional empirical correction is required for the fit shown. With $\gamma$ constrained to the Boyd--Kleinman value of $4.73\times10^{-5}~\mathrm{W}^{-1}$ (Table~\ref{tab:gammadelta}), reproducing the measured harmonic power requires $\delta=0.66\%$, corresponding to a cold finesse of 288 for the nominal input-coupler transmission and consistent with the measured value. The external blue-to-DUV conversion efficiency reaches 44\% when locked, against 18\% for the elliptical Brewster-cut cavity. On an internal basis, which is insensitive to the extraction geometry, the normal-incidence cavity still reaches 51\% against 31\%. The improvement reflects removal of the Brewster extraction loss together with the absence of a comparable empirical power-dependent conversion reduction under the optimized operating conditions.

This geometry delivers the highest DUV powers demonstrated in this work, and it sustains them without catastrophic coating failure or rapid grey-tracking, showing that AR-coated CZ-grown BBO can be operated near 230~nm under high CW DUV exposure when the bulk intensity and surface preparation are carefully controlled. Over long timescales, however, the AR coating eventually suffers visible laser-induced damage at the operating spot, at which point the beam is translated to a fresh location on the crystal; extending the coating lifetime remains the principal challenge for this geometry. The two geometries therefore serve different purposes: Brewster-cut BBO remains the robust default, tolerant of surface damage because there is no coating to lose, whereas normal-incidence AR-coated BBO is the higher-performance route, limited today by coating lifetime rather than by the crystal. The sustained output of all three cavity types is compared in Section~\ref{sec:longterm}.

\begin{table}[!htbp]
  \centering
  \caption{Internal, cavity, and external conversion efficiencies of the SHG cavities characterized in this work, quoted at their maximum measured values. The three conventions are defined in Section~\ref{sec:blueshg}. For the AR-coated crystals there is no Brewster extraction loss, so the internal and cavity efficiencies are equal to within the residual AR reflection. The external efficiency additionally accounts for the fraction of incident power coupled into the resonant mode, which is why it falls furthest below the cavity efficiency in locked operation, where the lower circulating power leaves the cavity further from impedance matching.}
  \label{tab:EfficiencyTable}
  \begin{tabular}{@{}llccc@{}}
    \toprule
    \textbf{SHG cavity} & \textbf{Operation} & \textbf{Internal} & \textbf{Cavity} & \textbf{External} \\
    \midrule
    Blue SHG (LBO, AR-coated)    & locked          & 94\% & 94\% & 88\% \\
    \addlinespace
    DUV Brewster-cut, spherical  & scan $=$ locked & 44\% & 34\% & 27\% \\
    \addlinespace
    DUV Brewster-cut, elliptical & scan            & 46\% & 35\% & 30\% \\
                                 & locked          & 31\% & 24\% & 18\% \\
    \addlinespace
    DUV AR-coated, spherical     & scan $=$ locked & 51\% & 51\% & 44\% \\
    \bottomrule
  \end{tabular}
\end{table}
\FloatBarrier

\subsection{Long-term behavior}\label{sec:longterm}
\begin{figure*}
    \centering
    \includegraphics[width=\textwidth]{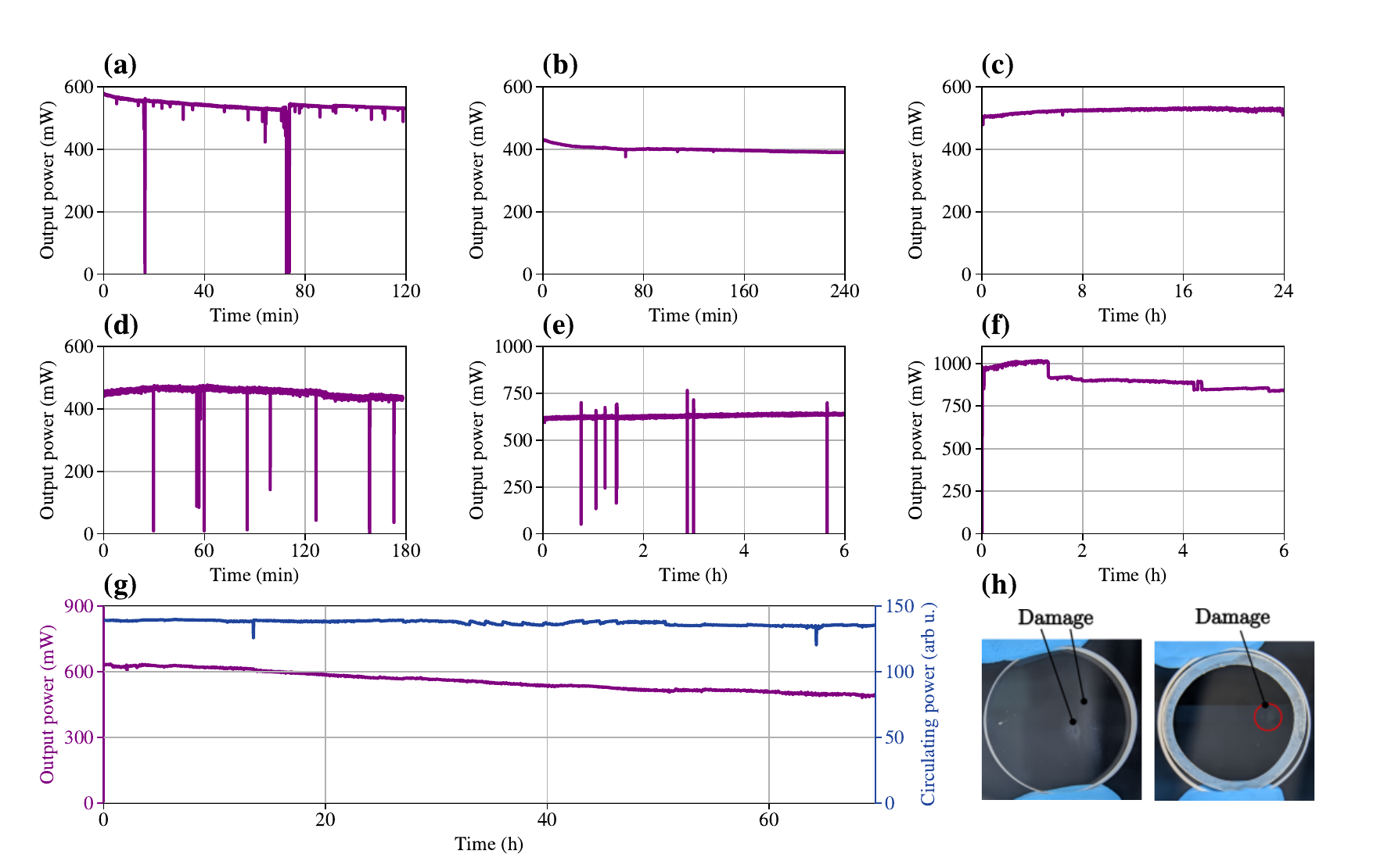}
    \caption{Long-term DUV output near 230~nm from six SHG enhancement cavities [panels (a)--(f)]. Elliptically focused, Brewster-cut BBO was used at (a) Imperial College London, (b) the University of Vienna, and (c) Agile Optic; a spherically focused, Brewster-cut crystal was used at (d) FHI Berlin; and spherically focused, normal-incidence AR-coated crystals were used at (e) FHI Berlin and (f) Agile Optic. The abrupt power dropouts correspond to cavity relocking events: thermal expansion of the cavity housing, caused by imperfect laboratory temperature stabilization, periodically drives the locking piezo beyond its operating range. Each trace ends because the measurement was stopped rather than because of degradation; all cavities continued to operate beyond the intervals shown. (g) Extended 70~h run of a spherical-focus Brewster-cut cavity at a single crystal position at Agile Optic, showing the DUV output power (left axis) and simultaneously monitored circulating blue power (right axis). The circulating power remains constant, indicating no degradation of the crystal or the intracavity optics. The slow decay of the delivered DUV power is instead caused by degradation of optics \textit{outside} the SHG cavity. (h) Photographs of the small defects subsequently found on the downstream dichroic mirror (left) and cavity output window (right). Rotating both components to fresh areas restored the initial DUV power; replacing them with a Brewster-angled window and the high-damage-threshold mirrors described in Section~\ref{sec:optics} provides a more durable solution.}
\label{fig:LongtermUVpower}
\end{figure*}

Having established the conversion efficiency of each geometry, we now compare how well they hold that performance over hours to days. Figure~\ref{fig:LongtermUVpower} collects the sustained DUV output of six cavities [panels~(a)--(f)], grouped by geometry. Of the elliptical-focus Brewster-cut cavities, the one at Imperial College London delivers up to about 580~mW, decaying slowly to $\sim$530~mW over two hours [panel~(a)]; a similar cavity at the University of Vienna delivers approximately 390--430~mW over four hours [panel~(b)]; and the cavity at Agile Optic holds 500--530~mW essentially without drift over 24 hours [panel~(c)]. The spherical-focus Brewster-cut cavity at FHI Berlin delivers 440--470~mW over several hours [panel~(d)]. The normal-incidence AR-coated cavities reach the highest sustained powers: about 620~mW, stable over six hours, at FHI Berlin [panel~(e)], and up to 1.0~W settling to 820--850~mW over six hours at Agile Optic [panel~(f)].

Several features are common to all six traces. The output can drift while the BBO and cavity reach thermal equilibrium, after which it remains comparatively stable; at the highest powers, residual thermal lensing can reduce the cavity enhancement, so the steady-state output may settle below the peak reached during a cavity scan. The occasional sharp dropouts are cavity relocking events (see the caption). Importantly, each trace ends because the measurement was stopped, not because of degradation or failure: the cavities continued to operate stably beyond the intervals shown.

Panel~(g) extends this to a continuous 70~h run of a spherical-focus Brewster-cut cavity held at a single position on the crystal, the longest uninterrupted measurement reported here. The delivered DUV power decays slowly over this interval, but the circulating blue power, monitored simultaneously on the same trace, remains constant. Since the circulating power is the quantity characterized by the round-trip loss of the enhancement cavity, its constancy shows that neither the BBO crystal nor the intracavity optics degraded, and that the decay originates downstream of the cavity. Inspection after the run confirmed this: small localized defects had formed on the cavity output window and on a dichroic mirror outside the cavity [panel~(h)]. Rotating both optics to move the beam onto fresh areas restored the initial DUV power. Replacing the flat output window with a Brewster-angled window and the dichroic with the high-damage-threshold mirrors described in Section~\ref{sec:optics} provides a more durable solution to this delivery-optics limitation. This carries a practical lesson for system design: at these wavelengths the delivery optics can fail before the nonlinear crystal does, so the circulating power is a more reliable witness of cavity health than the delivered power alone. We therefore recommend monitoring the fundamental circulating power alongside the delivered DUV power. Together, the two signals distinguish cavity degradation from delivery-optics degradation immediately and at no additional cost; without them, a slow decay in delivered power is easily misattributed to the nonlinear crystal, prompting an unnecessary crystal change.

\section{Optics for continuous-wave DUV operation}\label{sec:optics}
Stable operation near 230~nm requires not only efficient frequency conversion but also low-loss optics that tolerate prolonged CW DUV exposure. We therefore characterized cavity mirrors, polarizing beam splitters, and vacuum windows developed for this wavelength range in close collaboration with EKSMA Optics.

\subsection{Mirrors and cavity optics}\label{sec:mirrors}
Custom high-reflectance mirrors were manufactured by ion-beam sputtering, with ion-beam etching used where appropriate to improve the substrate surface before coating. Mirrors designed for $45^\circ$ incidence at 228.9~nm showed reflectances of 99.5\% for s-polarized light and 98.0\% for p-polarized light. Mirrors designed for $22.5^\circ$ incidence showed average reflectances of 99.2\% and 99.3\% for s- and p-polarized light, respectively, near 227.5~nm. Transmission measurements indicate residual absorption or scattering at the level of approximately 0.5\%.

We also tested high-reflectance coatings at 332~nm on superpolished Corning argon-fluoride (ArF)-grade fused-silica substrates. Although cleanroom measurements gave a transmission of 2~parts per million (ppm), resonant-cavity measurements under vacuum revealed total single-mirror losses of 446~ppm, indicating that transmission alone is not sufficient to qualify UV cavity optics: absorption and scattering must be measured directly.

\subsection{Polarizing beam-splitter plates}\label{sec:pbs}
\begin{figure}
    \centering
    \includegraphics[width=0.5\columnwidth]{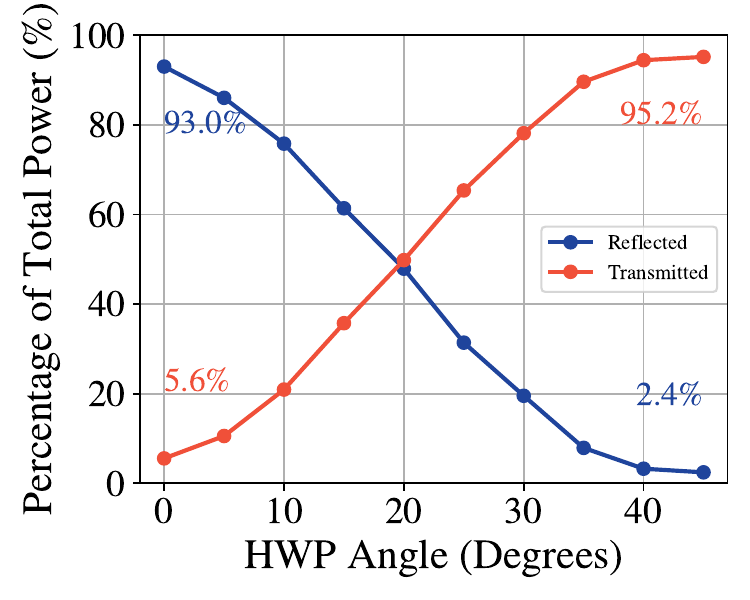}
    \caption{Measured performance of the thin-film DUV polarizing beam splitter near 230~nm: transmitted and reflected fractions of the total power as the input polarization is rotated with a half-wave plate (HWP). At the extremes, 95.2\% of p-polarized light is transmitted and 93.0\% of s-polarized light is reflected.}
    \label{fig:EksmaPBS}
\end{figure}

Efficient separation and recombination of DUV beams requires polarizing optics with high throughput and low absorption. Thin-film polarizing beam splitters were fabricated on 1~mm Corning ArF-grade fused-silica substrates and optimized for operation near 230~nm at $45^\circ$ incidence. The coating design uses aluminium-containing layers, surface etching, and post-process thermal annealing to reduce absorption and manage coating stress. As shown in Fig.~\ref{fig:EksmaPBS}, the plates provide 95\% p-polarization transmission and 93\% s-polarization reflection, with the summed transmitted and reflected powers accounting for 98--99\% of the incident power.

\subsection{Anti-reflection-coated windows under DUV operation}\label{sec:windows}
Vacuum-window degradation is a common limitation in DUV experiments, particularly in laser-cooling systems where beams must propagate through viewports into ultra-high vacuum. UV-induced dissociation of residual hydrocarbons can lead to contaminant adsorption on dielectric surfaces and a progressive reduction in transmission~\cite{Fosshaug2005SomeTools,Arnold2022UVContamination}. Motivated by the same surface-preparation considerations discussed for BBO, we tested whether ion-beam etching before coating improves the resistance of fused-silica windows to laser-induced contamination~\cite{Suratwala2015ChemistryGlass,Ray2023EnhancedLaserapplications,Zheng2012EffectSilica}.

\begin{figure*}[!htbp]
    \centering
\includegraphics[width=\textwidth]{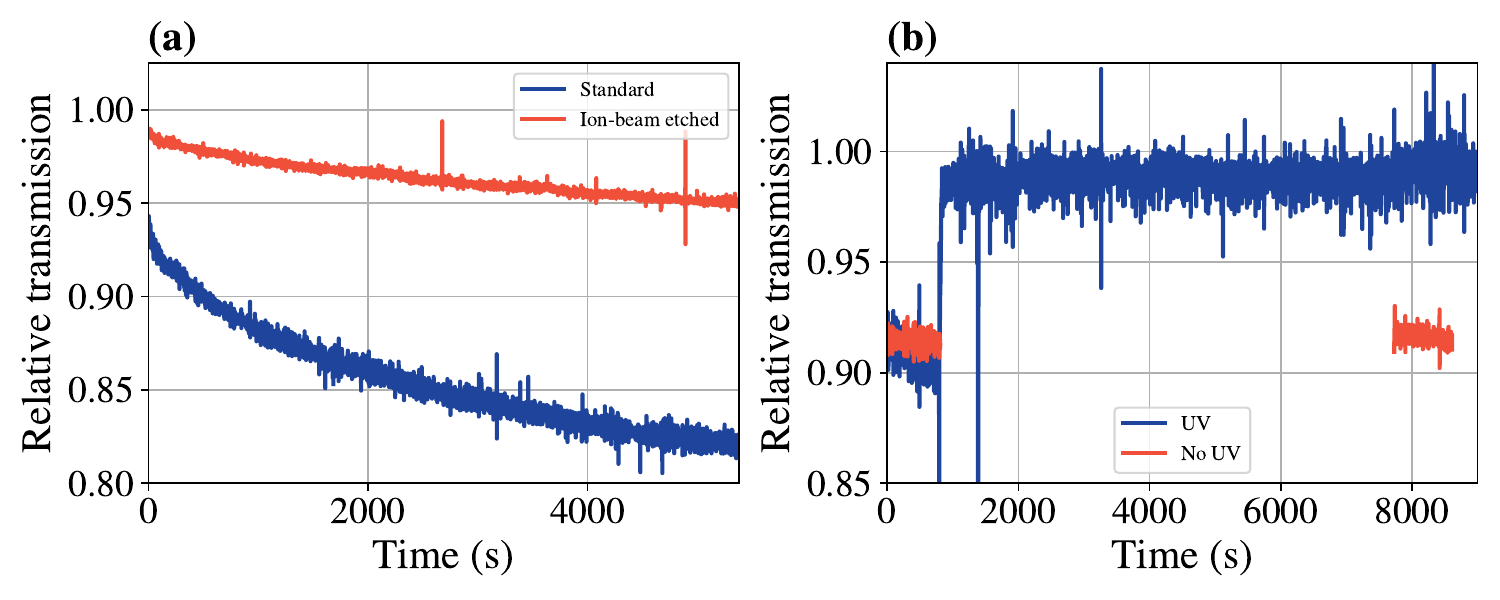}
    \caption{(a) Relative transmission through pairs of standard AR-coated windows and ion-beam-etched AR-coated windows under 227.5~nm irradiation at $11.5~\mathrm{W/cm^2}$. (b) Transmission recovery of ion-beam-etched windows during venting and subsequent pump-down. In one case (blue) the UV light remains on during venting and the transmission recovers fully; in the other (red) the UV light is turned back on only once the pressure has returned to $10^{-7}$~mbar, and no recovery is observed.}

    \label{fig:windows}
\end{figure*}

We compared two sets of fused-silica AR-coated windows: standard windows and windows whose surfaces were ion-beam etched before coating. DUV light was sent through a clean vacuum chamber maintained at $10^{-9}$~mbar, and the power transmitted through two windows was normalized to a pickoff before the chamber. At an intensity of $11.5~\mathrm{W/cm^2}$, the standard windows showed a 5\% transmission-decay time of 24~min, whereas the extrapolated decay time for the ion-beam-etched windows was 192~min, an eightfold improvement [Fig.~\ref{fig:windows}(a)]. The degradation rate is intensity dependent, and the decay is not single-exponential: an initial rapid loss is followed by a much slower residual decay once the surface reaches a quasi-steady state, so extrapolating the early slope alone underestimates the usable lifetime. For typical intensities of $1~\mathrm{W/cm^2}$ used in atomic and molecular MOT experiments, this residual rate is slow enough to allow experiments over years.

Figure~\ref{fig:windows}(b) shows that the transmission of a previously degraded spot recovers when the chamber is vented while the UV light remains on; venting without UV illumination produces no recovery. We attribute this cleaning effect to UV-generated reactive oxygen species produced from the oxygen admitted during venting; these species oxidize and remove adsorbed contaminants from the surface~\cite{Fosshaug2005SomeTools}.

\section{Applications: AlF molecular-beam spectroscopy and a Cd magneto-optical trap}\label{sec:applications}
As a system-level demonstration, we use the 227.5~nm laser to record laser-induced fluorescence on the $\mathrm{A}^1\Pi(v'=0)\leftarrow\mathrm{X}^1\Sigma^+(v''=0)$ transition of AlF. The DUV frequency is scanned over approximately 6~GHz by applying a voltage ramp to the piezo-mounted mirror of the 910~nm VECSEL, addressing the Q(1)--Q(6) lines. The laser intersects a cryogenic buffer-gas beam of AlF molecules in the detection region, 800~mm downstream from the cell.

\begin{figure*}
    \centering
    \includegraphics[width=\textwidth]{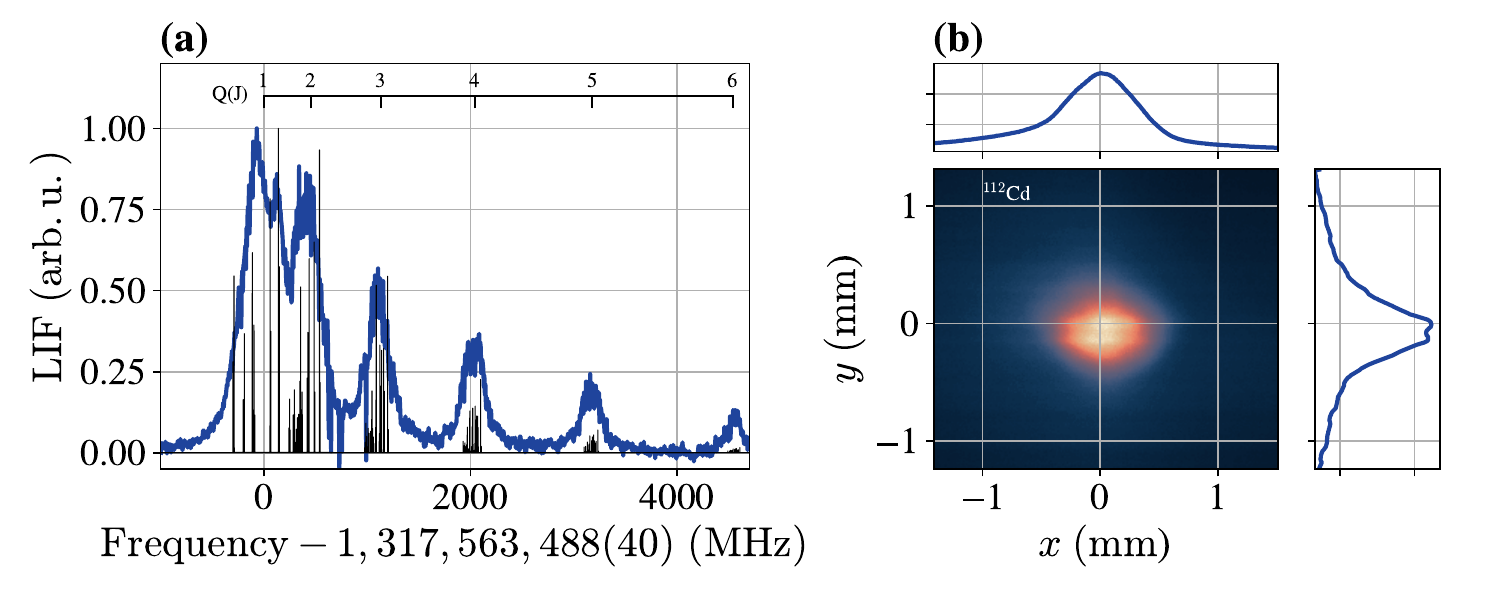}
    \caption{(a) Laser-induced fluorescence on the $\mathrm{A}^1\Pi (v'=0)\leftarrow \mathrm{X}^1\Sigma^+ (v''=0)$ transition of AlF, measured with a molecular beam from a cryogenic buffer-gas source and overlaid with Q-branch line positions simulated with PGOPHER. The DUV beam is retroreflected and intersects the molecular beam approximately perpendicular to its propagation direction. The laser power is 81~mW before the first input window; a 1.5~GHz scan of the VECSEL fundamental corresponds to a 6~GHz scan in the DUV. (b) Fluorescence image of a dispenser-loaded $^{112}$Cd magneto-optical trap containing $2\times10^5$ atoms, operated on the $^1S_0\rightarrow{}^1P_1$ transition at 228.9~nm. The trapping light is generated by a resonant BBO cavity of the type described in Section~\ref{sec:duv} and delivered using the mirrors, polarizing beam splitters, and AR-coated vacuum windows characterized in Section~\ref{sec:optics}; the trap therefore validates the complete DUV optical chain under continuous operation.}
    \label{fig:MOT+spectrum}
\end{figure*}

The measured spectrum is shown in Fig.~\ref{fig:MOT+spectrum}(a). The observed line positions agree well with a PGOPHER simulation~\cite{pgopher} using the spectroscopic constants from Ref.~\cite{Truppe2019}. The measured linewidth is dominated by the natural linewidth of the transition, $\Gamma/2\pi=84$~MHz, together with unresolved hyperfine structure in the ground and excited states. This measurement demonstrates that the VECSEL-based DUV source provides the tuning range and spectral stability required for molecular laser-cooling experiments.

The same architecture is used to laser-cool atomic cadmium, which offers a more stringent test because trapping requires sustained power delivered through the full optical chain rather than a single scanned beam. Figure~\ref{fig:MOT+spectrum}(b) shows a fluorescence image of a magneto-optical trap (MOT) of $^{112}$Cd operating on the $^1S_0\rightarrow{}^1P_1$ transition at 228.9~nm, loaded from a Cd dispenser and containing $2\times10^5$ atoms. Every DUV component developed in this work is used in this measurement: the trapping light is generated in an elliptical BBO cavity, then split, steered, and delivered into the vacuum chamber using the thin-film polarizing beam splitters of Section~\ref{sec:pbs}, the low-loss mirrors of Section~\ref{sec:mirrors}, and the ion-beam-etched AR-coated viewports of Section~\ref{sec:windows}. A DUV MOT is unforgiving of losses in this chain, and a stable trap is therefore good evidence that the optics maintain both throughput and wavefront quality under long-term CW DUV exposure.

The dispenser-loaded trap shown here is used for diagnostic purposes only. Recently, we used a similar 229~nm laser architecture to load large Cd MOTs from a cryogenic buffer-gas beam, capturing up to $1.1(2)\times10^7$ $^{112}$Cd atoms in 10~ms at a peak density of $2.5\times10^{11}$~cm$^{-3}$~\cite{PadillaCd2025}. Because Cd shares its cooling wavelength region, photon-scattering rate, and optics requirements with AlF, the Cd MOT also serves as a convenient day-to-day benchmark of the DUV laser systems used for molecular laser cooling~\cite{Padilla2025}.

\section{Discussion and outlook}\label{sec:discussion}

We have demonstrated compact VECSEL-based CW DUV laser systems near 230~nm, from the low-noise fundamental source to delivery at the experiment. External LBO cavities convert up to 94\% of the coupled NIR power and provide multi-watt blue output. The normal-incidence AR-coated BBO cavity reaches the highest DUV performance, producing up to 1.0~W and settling to 820--850~mW over six hours; this is, to our knowledge, the highest continuous-wave power reported from direct second-harmonic generation below 237~nm. AlF spectroscopy over a 6~GHz scan and a $^{112}$Cd magneto-optical trap validate the frequency control, beam quality, and complete DUV optical chain.

The BBO geometries have complementary trade-offs. Elliptical focusing reduces the peak fundamental intensity and produces a cleaner DUV mode, supporting long-term operation when combined with heating above 100$^\circ$C, a clean dry-air purge, and careful surface preparation. The theoretical (ideal) walk-off-plane beam quality improves from $M_x^2=2.47$ for spherical focusing to 1.09 for elliptical focusing. This may be an advantage for applications requiring a near-Gaussian mode, efficient beam delivery, or tight focusing. The spherical Brewster cavity produced up to 700~mW with a locked cavity efficiency of 34\%, but with a slightly fragmented output mode. The elliptical Brewster cavity reaches a cavity efficiency of 24\% when locked and requires an empirical power-dependent conversion correction, which likely also reflects power-dependent impedance matching. The absence of this correction in the spherical Brewster and AR-coated cavities suggests thermally induced mode-overlap loss in the more alignment-sensitive elliptical resonator. The AR-coated cavity avoids Brewster extraction loss and reaches 51\% cavity and 44\% external efficiency, although coating lifetime remains its principal limitation.

Further gains should come from more durable coatings and from slowly translating or dithering the crystal position. Distributing the optical dose over a larger area, rather than accumulating it at a single crystal or coating position, should extend the usable lifetime of each position while preserving continuous operation. The long-term data also show that delivery optics must be monitored separately from cavity performance.

Walk-off-compensated BBO offers a complementary route. In a bonded stack of equal-length plates with alternating optic-axis orientation, the transverse walk-off accumulated in one plate is reversed in the next, reducing the effective walk-off without sacrificing the total nonlinear interaction length~\cite{Friebe2006,Hara2012}. Such a device could permit a larger, more symmetric focus that combines high conversion efficiency with lower peak intensity and a cleaner DUV mode; the remaining question is whether the bonded interfaces withstand prolonged high-power CW DUV exposure. Across the UVQuanT program, 14 DUV cavities in 12 laser systems have operated at six laboratories across Europe over four years with only minimal servicing, indicating that the design principles reported here are robust and transferable.

\section*{Funding}
The UVQuanT collaboration (\url{https://www.uvquant.eu}) is funded by the Horizon Europe Framework Programme (101080164). S. Truppe acknowledges funding from the European Research Council (949119) and the Engineering and Physical Sciences Research Council (UKRI2226).

\section*{Acknowledgments}
We thank M. Arndt and S. Gerlich for useful discussions and G. Meijer for continuing support of UVQuanT.

\section*{Data availability}
The data underlying the results presented in this paper are openly available on
Zenodo in the UVQuanT community at
\url{https://zenodo.org/communities/101080164-uvquant/}.

\section*{Appendix: overview of UVQuanT systems}
\clearpage
\begin{sidewaystable}[p]
\centering
\caption{Overview of the collaboration's CW DUV laser systems and representative operating powers. Every DUV stage is a resonant BBO SHG cavity. ``Brewster'' and ``AR'' identify Brewster-cut and normal-incidence AR-coated crystals; ``elliptical'' and ``spherical'' identify the focusing geometry. Ext.: external enhancement cavity; IC: intracavity; typ.: routine operating power; max.: demonstrated maximum power.}
\label{tab:duv_systems_overview}
\scriptsize
\setlength{\tabcolsep}{3.5pt}
\renewcommand{\arraystretch}{1.18}
\begin{tabularx}{0.96\textheight}{@{}%
>{\hsize=0.85\hsize\linewidth=\hsize}Y
>{\hsize=0.92\hsize\linewidth=\hsize}Y
>{\hsize=0.50\hsize\linewidth=\hsize}Z
>{\hsize=1.70\hsize\linewidth=\hsize}Y
>{\hsize=0.80\hsize\linewidth=\hsize}Z
>{\hsize=0.90\hsize\linewidth=\hsize}Z
>{\hsize=1.22\hsize\linewidth=\hsize}Y
>{\hsize=0.92\hsize\linewidth=\hsize}Z@{}}
\toprule
\textbf{Location} & \textbf{Application} & \makecell{\textbf{DUV $\lambda$}\\\textbf{(nm)}} & \textbf{Source chain} & \makecell{\textbf{NIR}\\\textbf{(W)}} & \makecell{\textbf{Blue}\\\textbf{(W)}} & \textbf{DUV cavity} & \makecell{\textbf{DUV power}\\\textbf{(W)}} \\
\midrule
\multirow[t]{3}{=}{Agile Optic, Braunschweig} & \multirow[t]{3}{=}{SHG cavity development} & 230 & \multirow[t]{3}{=}{920~nm VECSEL;\\ ext.\ LBO $\to$ 460~nm} & \multirow[t]{3}{=}{4.2 output;\\ 3.7 delivered} & \multirow[t]{3}{=}{3.2 output;\\ 2.5 delivered} & Elliptical Brewster & 0.50 \\
\cmidrule(l){7-8}
 & & 230 & & & & Spherical Brewster & 0.70 \\
\cmidrule(l){7-8}
 & & 230 & & & & Spherical AR & 1.00 \\
\midrule
\multirow[t]{2}{=}{FHI Berlin} & \multirow[t]{2}{=}{Laser cooling of AlF} & 227.5 & Ti:sapphire (M~Squared); ext.\ LBO $\to$ 455~nm & 3.9 & 2.5 & Spherical Brewster & 0.44 \\
\cmidrule(l){3-8}
 & & 231.7 & Ti:sapphire (M~Squared); ext.\ LBO $\to$ 463.4~nm & 3.9 & 2.5 & Spherical AR & 0.60 \\
\midrule
\multirow[t]{4}{=}{Imperial College London} & \multirow[t]{4}{=}{Laser cooling of AlF and Cd} & 227.5 & Ti:sapphire (M~Squared); ext.\ LBO $\to$ 455~nm & 3.2 & 2.4 & Elliptical Brewster & 0.30 typ. \\
\cmidrule(l){3-8}
 & & 227.5 & 910~nm VECSEL; ext.\ LBO $\to$ 455~nm & 2.0 & $\leq1.4$ & Spherical Brewster & $\sim0.22$ typ. \\
\cmidrule(l){3-8}
 & & 227.5 & IC-doubled VECSEL $\to$ 455~nm & -- & $\leq1.5$ & Spherical Brewster & $\sim0.22$ typ. \\
\cmidrule(l){3-8}
 & & 231.7 & 926.8~nm VECSEL; ext.\ LBO $\to$ 463.4~nm & \makecell{5.4 output;\\ $\leq5.0$ delivered} & $>2.6$ reduced; 4.0 full & Elliptical Brewster & $\sim0.58$ max. \\
\midrule
\multirow[t]{2}{=}{University of Florence} & \multirow[t]{2}{=}{Laser cooling of Cd} & 229 & IC-doubled VECSEL $\to$ 458~nm & -- & 1.1 & Elliptical Brewster & $\sim0.07$ typ. \\
\cmidrule(l){3-8}
 & & 229 & 915~nm VECSEL; ext.\ LBO $\to$ 458~nm & 2.0 & 1.0 & Spherical Brewster & $\sim0.05$ max. \\
\midrule
\multirow[t]{2}{=}{University of Bonn} & \multirow[t]{2}{=}{Laser cooling of Zn} & 214 & Ti:sapphire (M~Squared); ext.\ LBO $\to$ 428~nm & 4.0 & 3.2 & Elliptical Brewster & $\sim0.15$ max. \\
\cmidrule(l){3-8}
 & & 214 & IC-doubled VECSEL $\to$ 428~nm & -- & 1.2 & Spherical Brewster & 0.10 \\
\midrule
University of Vienna & Nanocluster interferometry & 229 & Ti:sapphire (Sirah Matisse); ext.\ LBO $\to$ 458~nm & $>5.0$ & $>3.7$ output; $>3.1$ delivered & Elliptical Brewster & $\sim0.46$ max. \\
\bottomrule
\end{tabularx}

\end{sidewaystable}
\clearpage
\bibliographystyle{unsrtnat}
\bibliography{bibliography}

\end{document}